\documentclass[10pt,twocolumn,prl, aps, notitlepage, nofootinbib, showpacs, superscriptaddress,longbibliography]{revtex4-1}
\makeatletter
\def\l@subsubsection#1#2{}
\def\l@subsubsubsection#1#2{}
\makeatother
\usepackage[utf8]{inputenc}
\usepackage{graphicx,mathtools,xcolor,latexsym,rotating,soul}
\usepackage[T1]{fontenc}
\usepackage{tikz}
\usepackage{comment}
\usepackage{amsmath, amssymb, amsthm, amsfonts,breqn}
\usepackage{multirow}
\usetikzlibrary{decorations.pathmorphing}
\usepackage[colorlinks=true,citecolor=green,urlcolor=blue,linkcolor=blue]{hyperref}
\usepackage{pifont}
\usepackage{orcidlink}
\usepackage{gensymb}
\usepackage{enumitem}
\usepackage{mathrsfs}
\usepackage[scr=boondox,  
            cal=esstix]   
           {mathalpha}
\AddToHook{env/subequations/before}{\ifhmode\unskip\fi}
\allowdisplaybreaks

\makeatletter
\let\cat@comma@active\@empty
\makeatother
\begin{document}
\title{Relativistic and Dynamical Love}
\author{Abhishek Hegade K. R.}
\email{ah4278@princeton.edu}
\affiliation{Illinois Center for Advanced Studies of the Universe, Department of Physics, University of Illinois at Urbana-Champaign, Urbana, IL 61801, USA}
\author{K.J. Kwon \orcidlink{0000-0001-9802-362X}}
\email{jameskwon@ucsb.edu}
\affiliation{Department of Physics, University of California, Santa Barbara, CA 93106, USA}
\author{Tejaswi Venumadhav \orcidlink{0000-0002-1661-2138}}
\email{teja@ucsb.edu}
\affiliation{Department of Physics, University of California, Santa Barbara, CA 93106, USA}
\affiliation{International Centre for Theoretical Sciences, Tata Institute of Fundamental Research, Bangalore 560089, India}
\author{Hang Yu}
\email{hang.yu2@montana.edu}
\affiliation{eXtreme Gravity Institute, Department of Physics, Montana State University, Bozeman, MT 59717, USA}
\author{Nicolas Yunes}
\email{nyunes@illinois.edu}
\affiliation{Illinois Center for Advanced Studies of the Universe, Department of Physics, University of Illinois at Urbana-Champaign, Urbana, IL 61801, USA}

\begin{abstract}
Gravitational waves emitted in the late inspiral of binary neutron stars are affected by their tidal deformation. We study the tidal dynamics in full general relativity through matched-asymptotic expansions and show that the dynamical tidal response can be expanded in a set of modes. We further show that the mode amplitudes satisfy an effective, forced harmonic oscillator equation, which generalizes the overlap-integral formulation of Newtonian gravity.  Our relativistic treatment of dynamical tides will avoid systematic biases in future gravitational-wave parameter estimation, and can be generalized to model nonlinear tidal interactions and include the presence of elastic, electromagnetic, or dark matter fields interactions with nuclear matter.
\end{abstract}
\maketitle
\noindent\textbf{\textit{Introduction.}}
The shape of a neutron star immersed in an external gravitational environment differs from its equilibrium configuration due to tidal interactions.
The degree of tidal deformation depends on the strength of the tidal field and on the tidal response of the star.
Understanding how the tidal response affects the gravitational waves (GWs) emitted in a binary neutron star inspiral allows one to probe the internal structure of neutron stars using GW observations~\cite{Flanagan:2007ix,LIGOScientific:2018cki}.
Therefore, accurate modeling of the tidal response function is essential to prevent systematic errors in the inference of the equation of state~\cite{Chatziioannou_2020}.

The formalism to calculate the adiabatic tidal response of a neutron star was outlined in~\cite{Hinderer:2007mb,Binnington:2009bb,Damour:2009vw}. 
To use this formalism, the external tidal field experienced by the neutron star must change slowly, and there must be no resonant excitation of fluid modes inside the star. 
These assumptions fail in the presence of low-frequency $g$-mode~\cite{Lai:1993di,Yu:17a,Yu:17b, Andersson:18, Kwon:2024zyg,Kwon_2025} or inertial mode \cite{Ho:99, Flanagan:07, Xu:17, Poisson:20, Ma:21, Gupta:21} resonances, when there is tidal dissipation~\cite{Ripley:2023lsq,Ripley:2023qxo,HegadeKR:2024slr,Ghosh:2025wfx,Saketh:2024juq,Ghosh:2023vrx}, and during the late inspiral where the tidal field is dynamical~\cite{Steinhoff:2016rfi,Ma:20,Poisson:2020vap,Pitre:2023xsr,HegadeKR:2024agt, Yu:24a, Yu:25a}.

An elegant formalism to understand the \textit{dynamical} tidal response of a star in Newtonian gravity was begun in the late 20th century. 
This formalism relies on the fact that mode solutions of the linear perturbation equations 
(with a harmonic time-dependence $\sim e^{i \omega t}$)
of a non-rotating star are eigenfunctions of a self-adjoint operator~\cite{1964ApJ...139..664C,1965ApJ...142.1519C,gittins2025perturbationtheorypostnewtonianneutron,Smeyers-Book,yin2025postnewtonianapproachneutronstar} 
and hence, the spatial eigenfunctions form a complete basis.
When the tidal deformation is expanded in this basis, the coefficients obey the equations of a forced harmonic oscillator, with the `force' proportional to the overlap integral between the tidal field and the mode eigenfunction.
This simplification, in turn, enables the calculation of the tidal response function of the star as a mode-sum of harmonic oscillator response functions~\cite{Press-Teukolsky,Lai:1993di,Schenk_2001,Andersson_2020}. 

Extending the mode oscillation picture to full general relativity (GR) is hindered by multiple problems. One of these is the traditional restriction of calculations to isolated neutron stars and their loss of energy through gravitational wave emission with no-incoming-radiation boundary conditions~\cite{Kokkotas:1999bd}.  
These restrictions are problematic when considering neutron star binaries because (i) the neutron star is not isolated, and the gravitational field has both incoming and outgoing contributions due to the tidal field, and (ii) the radiative modes are quasi-normal in nature, and they cannot form a complete basis. Another important problem is that, in GR, one cannot easily separate the tidal field inside the star from the self-field the star generates, and thus, one cannot easily obtain a tidal force that excites oscillations in the star~\cite{Pitre:2023xsr}.
One way around these issues is to use matched asymptotic expansions combined with direct integration of the linearized perturbation equations~\cite{Poisson:2020vap,Pitre:2023xsr,HegadeKR:2024agt}.
Doing so, however, yields a purely numerical dynamical tidal response, which prevents the development of a physical picture of tidal oscillations. 

We here develop a physical picture of tidal dynamics in full GR in terms of modes by combining matched asymptotic expansions with resummation and analytic continuation. 
First, we prove that the mode solutions to the linearized perturbation equations in GR form a self-adjoint system if one matches the strong-field solution to a post-Newtonian (PN) solution (i.e.~a perturbative solution in small velocities and weak fields~\cite{Blanchet:2013haa}).
We then propose a novel treatment of the tidal excitation in GR by analytically continuing the tidal field inside the star in such a manner that when the tidal displacement is expanded in modes, the coefficients obey the equations of forced harmonic oscillators,
akin to the Newtonian approach.
Our approach provides a relativistic generalization of the overlap integral formulation, describing how the strong gravitational field of a neutron star contributes to its tidal deformation.
Moreover, our approach provides a relativistically-accurate, mode-sum framework to model the tidal excitations of a neutron star even in the late inspiral.

\vspace{0.5em}
\noindent\textbf{\textit{Background spacetime of a star.}}
Consider the background spacetime metric $g_{\mu\nu}$ of a non-rotating, spherically-symmetric, and static star in a general coordinate system\footnote{Latin indices $i, j, \ldots$ and Greek indices $\mu, \nu, \ldots$ represent purely spatial and spacetime coordinates respectively.} $(t,x^i)$ that admits a $3+1$ split and yields the infinitesimal line element
\begin{align}\label{eq:3+1-split}
    ds^2 &= g_{\mu \nu} dx^{\mu} dx^{\nu} = -N^2 dt^2 + \gamma_{ij} dx^i dx^j\,,
\end{align}
where henceforth we use the Einstein summation convention, the $(-,+,+,+)$ metric signature, and $G=1=c$. In Eq.~\eqref{eq:3+1-split}, $N$ is the lapse function and $\gamma_{ij}$ is the induced metric on $t=\mathrm{const}$ hypersurfaces. The quantity $\nabla_{\mu} t$ is a Killing covector of the background because of its stationarity. 
The stress-energy tensor is assumed to represent a perfect fluid, $T_{\mu \nu} = \varepsilon \, u_{\mu} u_{\nu} + p \, q_{\mu \nu}$, where $\varepsilon$ is the energy density, $p$ is the pressure, $\rho$ is the mass density, 
$u^{\mu}$ is the fluid four-velocity, and $q_{\mu\nu} \equiv u_{\mu} u_{\nu} + g_{\mu \nu}$ is a projection tensor.
The field equations of the background spacetime reduce to the Tolman-Oppenheimer-Volkoff equations in Schwarzschild coordinates.

\vspace{0.5em}
\noindent\textbf{\textit{Linear perturbations of the background.}}~The linearized perturbations arising from the tidal excitation of the background are governed by the linearized Einstein-Euler equations, which can be simplified using Lagrangian perturbation theory~\cite{Friedmann-rotating-stars,FN-book,FS_stability_rel} to
\begin{align}
\label{eq:basic-eqns-L-Pert-main-text-1}
    &\frac{\Delta \rho}{\rho} = -\frac{1}{2} q^{\mu \nu} \Delta g_{\mu\nu} \,,
    \frac{\Delta p}{p} = \Gamma \frac{\Delta \rho}{\rho}\,,
    \frac{\Delta \varepsilon}{\varepsilon + p} = \frac{\Delta \rho}{\rho}\,,\\
    &E_{\mu\nu}[\xi,h] \equiv \delta \left( G_{\mu \nu} - 8 \pi T_{\mu \nu}\right) = 0\,,\\
    &E^{\mu}[\xi,h] \equiv \delta \left[\nabla_{\nu} T^{\mu \nu} \right] = 0\,, 
\label{eq:basic-eqns-L-Pert-main-text-3}    
\end{align}
where $\delta$ and $\Delta = \delta + \mathcal{L}_{\xi}$ are the Euler and Lagrangian perturbation operators, respectively, $\xi^{\mu}$ is the Lagrangian displacement vector, $h_{\mu \nu}$ is the metric perturbation, and $\Gamma$ is the adiabatic index.
The explicit expressions of $E_{\mu\nu}$ and $E^{\mu}$ are provided in the Supp.~Mat.

In an elegant analysis, Friedman and Schutz~\cite{FS_stability_rel,Friedmann-rotating-stars} showed that for \textit{any} two, arbitrary, abstract vectors $y$ and $\hat{y}$, with components\footnote{Upper case Latin indices represent abstract components.} $\hat{y}_A = (\hat{\xi}_{\mu},\hat{h}_{\mu\nu})$ and $y_A = (\xi_{\mu},h_{\mu\nu})$, the following operator equation holds [Sec. 7.2 of~\cite{FN-book}]:
\begin{align}\label{eq:operator-Eqn-FS}
    \hat{\xi}_{\beta} E^{\beta}[y] + \frac{\hat{h}_{\alpha \beta} E^{\alpha \beta}[y] }{16 \pi} = - \mathscr{L}[\hat{y}, y] + \nabla_{\beta} \Theta^{\beta}[\hat{y}, y] \,,
\end{align}
where $\mathscr{L}[\hat{y}, y]$ and $\Theta^{\beta}[\hat{y}, y]$ are bilinear operators, and $\mathscr{L}[\hat{y}, y]$ is also symmetric, i.e.~$\mathscr{L}[\hat{y}, y]=\mathscr{L}[y,\hat{y}]$, but $\Theta$ is not necessarily so. 

We are here interested in seeing how this elegant formalism can be used to study the self-adjoint properties of the linearized equations. Without loss of generality, we use the gauge freedom in the definition of the Lagrangian displacement vector $\xi^{\mu}$ to set $\xi^{\mu} u_{\mu} = 0$.
Consider two abstract vectors ${y}$ and $\hat{y}$, where $y$ satisfies the linearized Einstein-Euler system, i.e., Eqs.~\eqref{eq:basic-eqns-L-Pert-main-text-1}--\eqref{eq:basic-eqns-L-Pert-main-text-3}, and $\hat{y}$ satisfies the linearized Hamiltonian and the momentum constraints (see Supp.~Mat.).
One can then show that the following operator form of the linearized Einstein-Euler system holds:
\begin{align}\label{eq:E-operator}
    \mathcal{E}[\hat{y}, y]
    \equiv 
    \hat{y}_{A} O_{0}^{AB} \partial_t^2 y_{B} + \hat{y}_{A} O_{1}^{AB} y_{B} + D_{\alpha}\mathcal{R}^{\alpha}[\hat{y},y] = 0\,.
\end{align}
Physically, the operator $O_{0}^{AB}$ describes the kinetic energy of the fluid and gravitational perturbations, and $O_{1}^{AB}$ corresponds to the potential energy.
The operators $O_{0,1}^{AB}$ are also symmetric bilinear operators, $\hat{y}_{A} O_{0,1}^{AB} y_{B} = \hat{y}_{B} O_{0,1}^{AB} y_{A} $.
The symmetry of the operator $\mathcal{R}^{\alpha}$, on the other hand, is not apparent without specifying a gauge for the metric perturbation and without specifying the boundary conditions.

If one is interested in the mode excitation of an isolated object, then the outgoing-radiation boundary condition is the correct condition to study. In that case, \cite{FS_stability_rel} showed that in the Fourier domain, $D_{\sigma} \mathcal{R}^{\sigma} \sim i \omega \, S$, where $\omega$ is a general (complex) Fourier frequency and $S$ is a symmetric positive operator. We see that, in this case, the $D_{\sigma} \mathcal{R}^{\sigma}$ operator encodes gravitational-wave dissipation in the system, leading to quasi-normal-like solutions. On the other hand, if one is interested in a relativistic star immersed in a tidal environment, then the boundary conditions cannot be purely outgoing, and they have to be modified.

\begin{figure}[htb]
    \centering
    \includegraphics[width=1\linewidth]{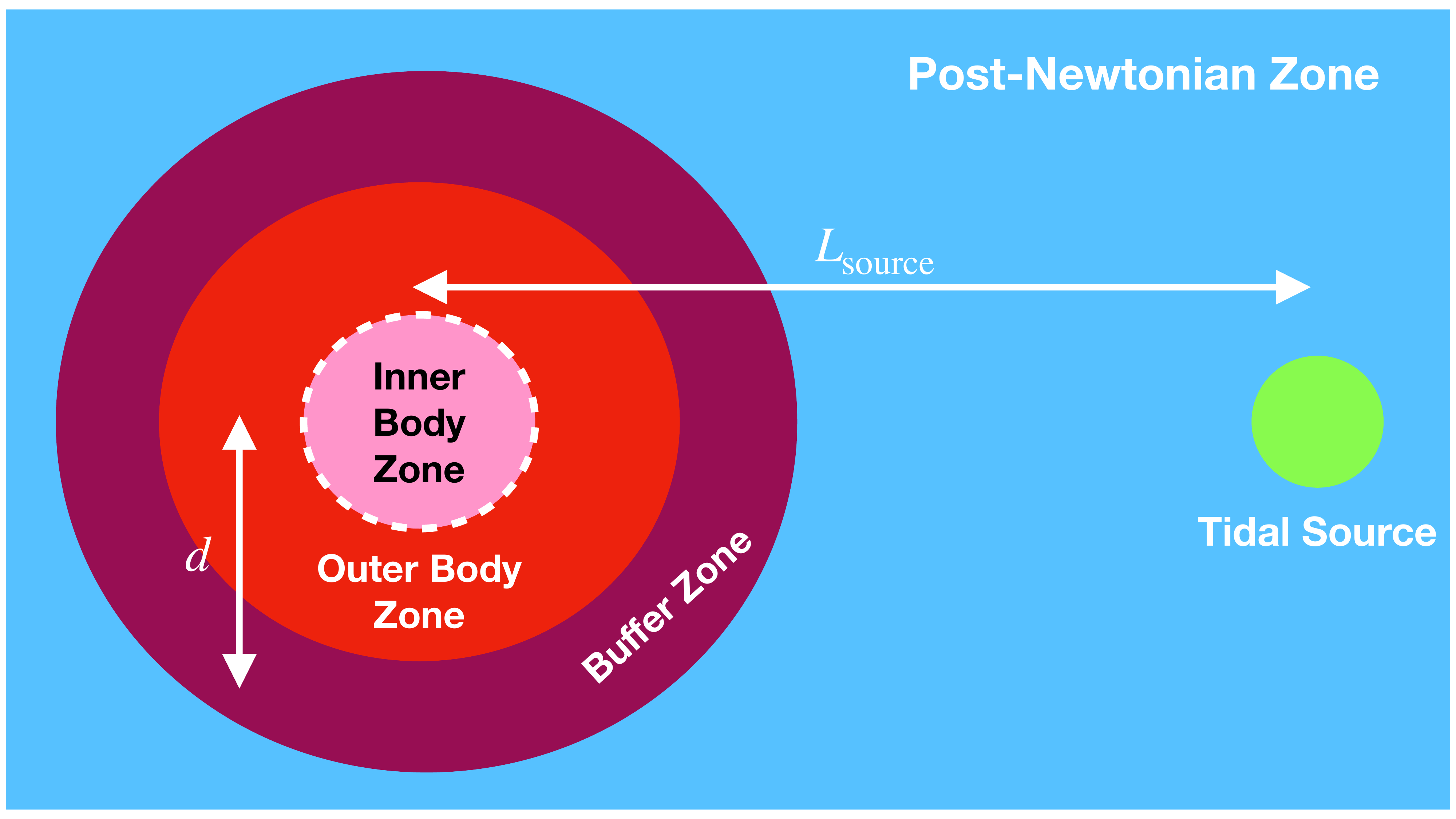}
    \caption{Cartoon (not to scale) of the near zone (full figure) in a tidally interacting system, consisting of a star (pink disk) of finite radius (dashed white circle), and a tidal source (green disk) located at a characteristic distance $L\sim L_{\mathrm{source}}$ from the center of mass of the star. We divide the near zone into different regions: inner and outer body zones, a buffer zone and a PN zone.
    The gravitational field in the inner (pink disk) and outer (red annulus) body zones is strong, while that in the PN zone (blue) is weak. The outer body and PN zone solutions are matched asymptotically in the buffer zone (purple annulus), which has a mean radius $d$ from the center of mass of the star.
    }
    \label{fig:zones}
\end{figure}

One approach to incorporate the presence of a tidal source is to use matched asymptotic expansions~\cite{DSX-I,DSX-II,DSX-III,Racine_2005,Poisson:2020vap,HegadeKR:2024agt}.
In this approach, the conservative dynamics of the body and the tidal environment are restricted to weak-field PN zone, and the dissipative (radiative) dynamics are analyzed by understanding the backreaction of GW emission on the system.
In the PN zone, the body and the external tidal source are assumed to be separated by a characteristic distance $L\sim L_{\mathrm{source}}$, as measured from the center of mass of the object, where $L_{\mathrm{source}}$ is large enough that the tidal interaction can be treated in PN theory.
To study how the PN environment influences the body, we separate the spacetime in the near zone into an inner body zone, an outer body zone, and the PN zone (see Fig.~\ref{fig:zones}).
In the inner and outer body zones, the gravitational field is strong, and we must solve the linearized Einstein-Euler system to understand the dynamics of the system.
However, the interaction between the body and the external source is weak in the PN zone, where the body appears as a `skeletonized' source with a multipolar structure~\cite{Poisson:2020vap}. 
To understand how the multipole moments depend on the internal properties of the body, the PN and body zone solutions must be asymptotically matched in a buffer zone, where both the strong-field internal and the PN solutions are valid. 

\vspace{0.5em}\noindent
\textbf{\textit{Mode expansion of the perturbations.}}
We now prove that, if the strong field solution of the linearized perturbation equation matches asymptotically to a PN solution that describes the interaction between the object and the tidal source in the PN zone, then the linearized Einstein-Euler system is self-adjoint.
This result is important because it allows us to expand these perturbations in a mode basis, and then to show that the tidal perturbations of the star can be understood as a sum of harmonic oscillators that respond to an external force.

Suppose that the coordinate system outside the star is $(t, R \sin(\theta) \cos(\phi), R \sin(\theta) \sin(\phi), R \cos(\phi))$ where $R = r-M_{\star}$, $M_{\star}$ is the mass of the star and $(t,r,\theta,\phi)$ are Schwarzschild coordinates. 
Let $h_{\mu \nu}$ satisfy the harmonic gauge condition outside the star, $\nabla_{\mu} \bar{h}^{\mu \nu} = 0$, where $\bar{h}_{\mu\nu} = h_{\mu \nu} - \frac{1}{2} g_{\mu\nu} h^{\alpha}_{\alpha}$ is the trace-reversed metric perturbation. 
Assume that the metric perturbation asymptotically matches the 1PN solution (see Supp.~Mat.~for explicit expressions) in the buffer zone, which is assumed to be located at a mean radius of $R =  d $.
Then, the mode solutions for a particular set of spherical harmonic modes $(\ell,m)$ can be shown to satisfy the following operator equation, after taking the Fourier transform:
\begin{align}\label{eq:mode-eqn-harmonic-main}
    &\omega^2 \int \sqrt{\gamma} d^3 x\left[ \hat{y}_{A} O_{0,\mathrm{harm}}^{AB} y_{B}\right]
    =
    \int \sqrt{\gamma} d^3 x \left[\hat{y}_{A} O_{1,\mathrm{harm}}^{AB} y_{B}\right]
    \nonumber\\
    &\hspace{1cm}
    +
    \left[\frac{2 \pi  \tilde{I}_{\ell m} \hat{\tilde{I}}_{\ell m} (\ell-2) M}{c^2 (2 \ell+1)^2 d^{2 (\ell+1)}}+ \mathcal{O}(c^{-4}) \right]\,,
\end{align}
where we have reintroduced factors of $c$ to highlight the PN nature of the solution in the buffer zone, $\gamma$ is the determinant of the induced metric, $\omega$ is the Fourier frequency, and $\tilde{I}_{\ell m}(\omega)$ and $\hat{\tilde{I}}_{\ell m}(\omega)$ are the Fourier transforms of the multipole moments of the solutions $y$ and $\hat{y}$, respectively.
Equation~\eqref{eq:mode-eqn-harmonic-main} holds for any general vector $\hat{y}$ that satisfies the Hamiltonian and momentum constraints.
Since both operators on the left- and right-hand sides of Eq.~\eqref{eq:mode-eqn-harmonic-main} are symmetric operators, we conclude that the mode solutions are eigenfunctions of a self-adjoint operator.
We shall assume that these modes form a complete set of solutions as one does in Newtonian and PN theory~\cite{1964ApJ...139..664C,1965ApJ...142.1519C,gittins2025perturbationtheorypostnewtonianneutron,yin2025postnewtonianapproachneutronstar}.
One can prove a similar statement in Regge-Wheeler gauge, but we leave this to another publication~\cite{future-work-in-prep}.

We note that the self-adjoint operator constructed in Eq.~\eqref{eq:mode-eqn-harmonic-main} is based on matched-asymptotic expansions, i.e., in our approach, the gravitational field in the inner body zone can be arbitrarily strong.
The only restriction of our approach is the existence of a buffer zone where a 1 PN metric (see Supp. Mat. for a discussion of extension to higher PN orders) can be used to describe the tidal interactions. This restriction is valid during the inspiral of binary neutron stars.
In contrast to this work, recent approaches are based on \textit{global} PN approximations~\cite{gittins2025perturbationtheorypostnewtonianneutron,yin2025postnewtonianapproachneutronstar}, where the dynamics of the system in the near-zone is required to be strictly a PN solution. Using such global 1 PN approximations in the inner body zone can lead to large systematic errors due to the strong gravitational field of a neutron star. 

\vspace{0.5em}
\noindent
\textbf{\textit{Tidal excitation.}}
In Newtonian gravity, there is a formalism that allows one to understand the tidal excitation of fluid motion inside a star by decomposing the fluid perturbation in terms of the mode solutions of the homogeneous problem~\cite{Lai:1993di,Press-Teukolsky,Schenk_2001,Andersson_2020}.
We now extend this formalism to full GR using the operator framework discussed above. For simplicity, we  work in harmonic gauge, but the technique also works in RW gauge~\cite{future-work-in-prep}.

Let $\textsf{h}_{\mu \nu}$ denote the metric perturbation induced by the tidal field of the external environment.
Outside the star we can decompose the gravitational perturbation as $\textsf{h}_{\mu \nu}^{\mathrm{ext}} = h_{\mu \nu}^{\mathrm{mult}} + h_{\mu \nu}^{\mathrm{\mathrm{T, ext}}}$, where the multipolar $h_{\mu \nu}^{\mathrm{mult}}$ and the tidal $h_{\mu \nu}^{\mathrm{\mathrm{T},\mathrm{ext}}}$ pieces asymptotically match the PN solutions characterized by multipoles $I_{\ell m}(t)$ and tidal moments $d_{\ell m}(t)$, respectively.
See the Supp.~Mat.~for explicit expressions of the PN solutions.

Let us now extend this decomposition into multipolar and tidal pieces to the interior of the star. 
No \textit{unique} decomposition of the solution into multipolar and tidal pieces exists inside the star; however, as we demonstrate below, there are a set of \textit{natural} decompositions that allows us to view the tidal piece as a force density that sources oscillations inside the star.
The linearized Einstein-Euler system inside the star is given by
\begin{align}
\label{eq:tidal-perturbation-v1}
    &E_{\mu \nu}[\xi,\textsf{h}] \equiv E_{\mu \nu}[\xi,h] + E_{\mu \nu}^{\mathrm{grav}}[h^{\mathrm{T}}] = 0\,\\
    &E_{\mu}[\xi,\textsf{h}] \equiv E_{\mu}[\xi,h] + E_{\mu}^{\mathrm{grav}}[h^{\mathrm{T}}] = 0\,,
\label{eq:tidal-perturbation-v2}
\end{align}
where $h_{\mu \nu}$ and $h^{\mathrm{T}}_{\mu \nu}$ are an \textit{arbitrary} split of the metric perturbations inside star. 
The operators $E^{\alpha \beta}_{\mathrm{grav}}[h^T] \equiv E^{\alpha \beta}[\xi=0,h^T]$ and $E_{\mu}^{\mathrm{grav}}[h^T] \equiv E_{\mu}[\xi=0,h^T]$ are purely ``gravitational operators'' that do not excite any fluid displacement. 
Now, consider \textit{any} set of functions $h^{\mathrm{T}}_{\mu \nu}$ that match continuously and differentiably at the surface of the star to $h_{\mu \nu}^{\mathrm{\mathrm{T, ext}}}$ and satisfy the ``tidal Hamiltonian'' and ``tidal momentum constraints'', $E_{\mu \nu}^{\mathrm{grav}} [h^T] u^{\mu} = 0$. 
These constraints provide a set of differential equations which can be combined with the tidal boundary conditions to solve for, and hence extend, certain components of the tidal field inside the star.
Using these solutions, the linearized equations of motion [Eqs.~\eqref{eq:tidal-perturbation-v1} and \eqref{eq:tidal-perturbation-v2}] simplify to
\begin{align}
\label{eq:force-split-1}
    &E_{\mu \nu}[\xi,h] = -E_{\mu \nu}^{\mathrm{grav}}[h^{\mathrm{T}}] \equiv F_{\mu\nu}[h^T] \,,\\
    &E_{\mu}[\xi,h] = - E_{\mu}^{\mathrm{grav}}[h^T] \equiv F_{\mu}[h^T]\,.
\label{eq:force-split-2}
\end{align}
A physical meaning of the above equations then naturally arises: one can view the tidal metric perturbation as a source that excites fluid and gravitational  oscillations inside the star. 

The force density $F^{\mu}$ can be schematically split as a linear combination of $F^{\mu}_{\mathrm{dens,T}}$, $F^{\mu}_{\mathrm{T,grad.\,p}}$ and $F^{\mu}_{\mathrm{dens,vec}}$. The first piece, $F^{\mu}_{\mathrm{dens,T}}$, is the force density that arises from the coupling between the gradients of the tidal field and the energy density of the star. The second piece, $F^{\mu}_{\mathrm{T,grad.\,p}}$, arises due to the coupling between the tidal field and the pressure gradients inside the star. The third piece, $F^{\mu}_{\mathrm{dens,vec}}$, describes the coupling of the energy density and the vector potential of the tidal field.
In the Newtonian limit, only $F^{\mu}_{\mathrm{dens,T}}$ survives, whereas in GR, all forms of energy and gravitational gradients contribute to the force.
The tidal stress-energy tensor $F_{\mu\nu}$ is a purely spatial tensor ($F_{\mu \nu} u^{\nu} = 0$) that couples the dynamical gravitational degrees of freedom inside the star with the tidal field.

With the linearized metric perturbation equation in hand, we can then combine them to obtain an operator form of the equations for the ``multipolar piece'' $y$ inside the star. Following the same steps that lead to Eq.~\eqref{eq:E-operator} we then find
\begin{align}\label{eq:general-force-eqn-operator}
    \mathcal{E}_{\mathrm{harm}} &= 
    \hat{y}_{A} O_{0,\mathrm{harm}}^{AB} \partial_t^2 y_{B} 
    + 
    \hat{y}_{A} O_{1,\mathrm{harm}}^{AB} y_{B} 
    + 
    D_{\sigma} \mathcal{R}_{\mathrm{harm}}^{\sigma}[\hat{y},y]
    \nonumber\\
    &= 
    N \hat{\xi}_{\beta} F^{\beta}
    +
    \frac{N\hat{h}_{\alpha \beta} F^{\alpha \beta}}{16 \pi}
    \,,
\end{align}
in harmonic gauge. The formal solution to the above equations can be found by decomposing the perturbations $y_{B}$ in terms of the eigenfunction of the homogeneous equation, $y_{B} = \sum_{s} a_{s}(t) \, y_{B,s}$ where $s$ denotes the mode solution with frequency $\omega_{s}$. 
The functions $y_{B,s}$ satisfy $\hat{y}_{A} O_{0,\mathrm{harm}}^{AB} y_{B,s} \omega_s^2 = \hat{y}_{A} O_{1,\mathrm{harm}}^{AB} y_{B,s} + D_{\sigma} \mathcal{R}_{\mathrm{harm}}^{\sigma}[\hat{y},y_{B,s}]$.
Substituting the decomposition into Eq.~\eqref{eq:general-force-eqn-operator} and using the mode solution and simplifying, we obtain
\begin{align}\label{eq:forced-oscillator-v1}
    \!\!\sum_{s} \hat{y}_{A} {y}_{B,s} O_{0,\mathrm{harm}}^{AB} \left[
    \!
    \frac{d^2 a_{s}}{dt^2}  
    +\omega^2_{s} a_{s} 
    \!
    \right]
    \!\!=
    \!\!
    N\hat{\xi}_{\beta} F^{\beta}
    \!\!+\!\!
    \frac{N\hat{h}_{\alpha \beta} F^{\alpha \beta}}{16 \pi}.
\end{align}
We can normalize the mode solutions as 
\begin{align}\label{eq:normalization-condition-star}
    \int d^3 x \sqrt{\gamma} \, y_{A,s}^{*} {y}_{B,p} O_{0,\mathrm{harm}}^{AB}= \frac{R_{\mathrm{\star}} M_{\star}^2}{(R_{\mathrm{\star}}\omega_{s})^2} \delta_{sp} \mathscr{N}_{s}
\end{align}
where $M_{\star}$ is the stellar mass and $R_{\star}$ the stellar radius, and $\mathscr{N}_{s}$ is a dimensionless normalization factor. Integrating Eq.~\eqref{eq:forced-oscillator-v1} against $\hat{y}_{A} = y^{*}_{A,p}$, we obtain a driven, harmonic oscillator equation, 
\begin{align}
    \frac{d^2 a_{s}}{dt^2} + \omega^2_{s} a_{s} = {\cal{F}}_s\,,
    \label{eq:sec_order_sys}
\end{align}
where we have defined the effective driving force
\begin{align}
\label{eq:def-driving-force}
    {\cal{F}}_s &\equiv \frac{(R_{\mathrm{\star}}\omega_{s})^2}{\mathscr{N}_s(R_{\mathrm{\star}} M_{\star}^2)}\int d^3 x N \sqrt{\gamma} \, \xi^{*}_{{\beta},s} F^{\beta} 
    \nonumber\\
    &+
    \frac{(R_{\mathrm{\star}}\omega_{s})^2}{\mathscr{N}_s(R_{\mathrm{\star} } 16 \pi M_{\star}^2)}\int d^3 x N \sqrt{\gamma} \, h^{*}_{{\alpha \beta},s} F^{\alpha \beta} 
    \,.
\end{align}
The tidal source induces forced oscillations inside the star, similar to those in Newtonian gravity, except that now the driving force contains additional relativistic contributions coming from the force densities $F^{\mu}_{\rm T,grad. p}$, $F^{\mu}_{\rm dens, vec}$ and the tidal stress-energy tensor $F_{\mu \nu}$.

\vspace{0.5em}
\noindent
\textbf{\textit{The tidal response function.}}
We now assume that all perturbed quantities have harmonic time dependence.
The decompositions of the mode amplitude, the multipole moment, the tidal moment, the force density and tidal stress-energy tensor are given by  $a_{s}(t) = \tilde{a}_{s}(\omega) e^{i \omega t}$,  $I_{\ell m} = \tilde{I}_{\ell m}(\omega) e^{i \omega t}$,  $d_{\ell m}(t) = \tilde{d}_{\ell m}(\omega) e^{i \omega t}$,  $F_{\mu} = e^{i\omega t} \Tilde{F}_{\mu}(\omega)$,  $F_{\mu \nu} = e^{i\omega t} \Tilde{F}_{\mu \nu}(\omega)$ respectively. 
We also define the overlap $I_{s}$ so that it satisfies
\begin{align}\label{eq:Is-def}
    &\frac{4 \pi}{2 \ell + 1}\tilde{d}_{\ell m} I_{s}  
    \equiv 
    \frac{1}{M_{\star} R_{\star}^{{\ell}}}\int d^3 x \sqrt{\gamma} N \xi^{*}_{{\beta},s} \Tilde{F}^{\beta}
    \nonumber\\
    &+
    \frac{1}{M_{\star} R_{\star}^{{\ell}} 16 \pi}\int d^3 x N \sqrt{\gamma} \, h^{*}_{{\alpha \beta},s} \Tilde{F}{}^{\alpha \beta} 
    \,.
\end{align}
The particular solution to the driven, harmonic oscillator differential equation in the frequency domain is then 
\begin{align}\label{eq:sol-a-freq}
    \tilde{a}_s(\omega) = \frac{4 \pi }{2 \ell + 1} \; \frac{\tilde{d}_{\ell m}}{\mathscr{N}_s \, M_{\star}/R_{\star}^{\ell +1}} \, \frac{I_{s}}{1 - (\omega/\omega_s)^2}\,.
\end{align}

We now use the above result to obtain an expression for the tidal response function of the star $\tilde{K}_{\ell m}(\omega)$, that relates the multipole moment of the star to the tidal moment of the external spacetime via
\begin{align}
    \tilde{I}_{\ell m}(\omega) = 2 \tilde{K}_{\ell m}(\omega) R_{\star}^{2\ell + 1} \tilde{d}_{\ell m}(\omega) \,.
\end{align}
Note that there can also be homogeneous contributions to the response that describes the interaction of tidal field with the past resonance history of the star~\cite{Lai:1993di, Yu:24a}. Although one can describe this effect in terms of Eq. (\ref{eq:sec_order_sys}), we leave a detailed analysis to future work~\cite{future-work-in-prep}.

As we demonstrate in the Supp.~Mat., the integration of the Hamiltonian constraint for $h_{\mu\nu}$ and $h^{T}_{\mu\nu}$ provides the formula for the tidal response function
\begin{align}\label{eq:dynamical-tidal-response-func-GR}
    \tilde{K}_{\ell m}(\omega) &=  
    \frac{1}{\mathcal{Q}(\ell, z_0, \omega) }
    \sum_{s} \frac{2\pi}{2\ell+1} \frac{I_{s} \mathcal{G}_{s}}{\left[1 - (\omega/\omega_s)^2\right] \mathscr{N}_s}\,,
\end{align}
where $z_{0} \equiv 1-2M_{\star}/(R_{\star}+M_{\star})$.
The quantities $I_{s}$ and $\mathcal{G}_{s}$ are the relativistic generalization of the dimensionless tidal overlap integrals of Newtonian gravity. 
In fact, in the Newtonian limit, we have that $\mathcal{Q}_{\mathrm{Newt}} = 1$ and
\begin{align}
    \label{eq:Newt-limit-overlap-1}
    &I_{s,\mathrm{Newt}} = \mathcal{G}_{s,\mathrm{Newt}} = \frac{1}{M R_{\star}^{\ell}}\int d^3 x \, \rho \, \xi^{*}_{j,s}  \partial^j (r^{\ell} Y_{\ell m} )\,,\\
    &\tilde{K}_{\ell m,\mathrm{Newt}} =  
    \frac{2\pi}{2\ell + 1}
    \sum_{s} \frac{(I_{s,\mathrm{Newt}})^2}{\left[1 - (\omega/\omega_s)^2\right] \mathscr{N}_s}\,.
    \label{eq:Newt-limit-overlap-2}
\end{align}
where $Y_{\ell m}$ are spherical harmonics and $\rho$ is the baryon mass density.
In GR, $I_{s}$ and $\mathcal{G}_{s}$ can be split schematically into (see Supp.~Mat.~for explicit expressions)
\begin{align}
    &I_{s} = I_{s}^{\mathrm{dens,T}} + I_{s}^{\mathrm{T,grad.\,p}} + I_{s}^{\mathrm{dens,vec}} + I_{s}^{\mathrm{TS}}\,,\\
    &\mathcal{G}_{s} = \mathcal{G}_{s}^{\mathrm{dens,T}} + \mathcal{G}_{s}^{\mathrm{T,grad.\,p}} + \mathcal{G}_{s}^{\mathrm{grav,T}}\,.
\end{align}
The quantities $I_{s}^{\mathrm{dens,T}}$ and $\mathcal{G}_{s}^{\mathrm{dens,T}}$ are the dimensionless tidal overlap integrals that arise from the coupling between the gradients of the tidal field and the energy density of the star, and they are the only terms that survive in the Newtonian limit [Eq.~\eqref{eq:Newt-limit-overlap-1}]. The contributions $I_{s}^{\mathrm{T,grad.\,p}}$ and $\mathcal{G}_{s}^{\mathrm{T,grad.\,p}}$ arise from the coupling of the tidal field with the pressure gradients inside in the star. Finally, $I_{s}^{\mathrm{dens,vec}}$ is the component of the tidal overlap that couples the tidal perturbation to the gravitational vector potential of the star, $I_{s}^{\mathrm{TS}}$ arises from the tidal stress-energy contribution, and $\mathcal{G}_{s}^{\mathrm{grav,T}}$ is a purely gravitational contribution that couples the gravitational perturbation to the tidal field. 

\vspace{0.5em}
\noindent
\textbf{\textit{Discussion.}}
We have developed a relativistic formalism to quantify the impact of dynamical tidal effects in full GR. 
Using PN boundary conditions in the buffer zone, we established that the mode amplitudes behave as an effective harmonic oscillator and that the tidal response in full GR can be obtained using overlap integrals. These results provide a rigorous and physical basis for phenomenological models of tidal interactions used in effective one-body models~\cite{Hinderer:2016eia,Steinhoff:2016rfi}.
This formalism will become an essential tool for studying the oscillations of neutron stars in relativistic tidal environments, such as in the late inspiral of binary neutron stars. For example, this formalism enables us to robustly quantify the importance of relativistic corrections to dynamical tidal excitations in the gravitational waves emitted before the merger of binary neutron stars.
Other applications include the study of $g$-mode excitations in GR, understanding their impact on the gravitational waves, formulating the general relativistic version of the tidal dissipation problem studied in~\cite{Lai:1993di}, and understanding resonance locking of neutron stars~\cite{Kwon:2024zyg,Kwon_2025}.
This formalism also lends itself to clean extensions applicable to several interesting cases, such as the tidal response of slowly-rotating neutron stars, non-linear tidal excitations~\cite{Weinberg_2012,Yu_2022,pitre2025impactnonlinearitiesrelativisticdynamical}, and dynamical tides in the presence of extra (scalar or vectorial) degrees of freedom, such as elastic fields, electromagnetic fields, or dark matter fields~\cite{FS_stability_rel}.
\begin{acknowledgments}
\vspace{0.5em}
\noindent{{\bf{\em Acknowledgments.}}}
We thank Nils Andersson, Eric Poisson, Huan Yang, Hao-Jui Kuan, Kostas Kokkotas and Arthur Suvorov for discussions on relativistic tidal interactions and for their detailed comments on the paper. 
A.H. and N.Y. acknowledge support from the Simons Foundation through Award
No. 896696, the NSF through Grant No. PHY-2207650
and NASA through Grant No. 80NSSC22K0806. 
T.V. and K.J.K. acknowledge support from NSF grants 2012086 and 2309360, the Alfred P. Sloan Foundation through grant number FG-2023-20470, and the BSF through award number 2022136.
H.Y. is supported by NSF grant No. PHY-2308415 and Montana NASA EPSCoR Research Infrastructure Development under award No. 80NSSC22M0042. 
\end{acknowledgments}
\bibliography{ref}
\clearpage
\section*{SUPPLEMENTARY MATERIAL}
\section{Outline of the proof of self-adjointness of the operator equation}
\subsection{Background Einstein and stress energy conservation equations}
The TOV equations in $3+1$ form are given by
\begin{subequations}\label{eq:TOV-general-equations}
\begin{align}
    &\,{}{}{}{}{}^{(3)}R - 16 \pi \varepsilon = 0\,,\\
    &\,{}{}{}{}{}^{(3)}G_{\rho \sigma } N - 8 \pi \gamma_{\rho \sigma } p N + \gamma_{\rho \sigma } D_{\alpha }D^{\alpha }N -  D_{\sigma }D_{\rho }N = 0\,,\\
    &D_{\sigma }p + \frac{h \rho D_{\sigma }N}{N} = 0\,,
\end{align}
\end{subequations}
where the $(3)$ superscript denotes quantities intrinsic to the spatial metric $\gamma_{\mu\nu}$.
If we use Schwarzschild coordinates $(t,r,\theta,\phi)$, Eq.~\eqref{eq:TOV-general-equations} reduces to the familiar TOV equations [see Eqs. (6.2)-(6.4) of~\cite{FN-book}].

One useful consequence of Eq.~\eqref{eq:3+1-split} of the main text and the fact that $\nabla^{\mu} t$ is a Killing vector is that the extrinsic curvature is equal to zero,
\begin{align}
    K_{\alpha \beta} = 0\,.
\end{align}
We shall use this fact when simplifying many expressions below.
\subsection{Linearized Einstein-Euler system}
The basic equations of Lagrangian fluid perturbation theory were provided in Eqs.~\eqref{eq:basic-eqns-L-Pert-main-text-1}-\eqref{eq:basic-eqns-L-Pert-main-text-3} of the main text. We now list other expressions explicitly, which were mentioned in the main text.
The linearized Einstein equation $E_{\mu\nu}$ and the stress-energy conservation equations $E^{\mu}$ are
\begin{widetext}
\begin{subequations}\label{eq:linearized-Einstein-Euler-system}
\begin{align}
    E^{\alpha \beta}[\xi,h] &= -\frac{1}{2} \epsilon^{\alpha \gamma \epsilon \zeta} \epsilon^{\beta \delta \eta}{}_{\zeta} \nabla_{\gamma} \nabla_{\delta} h_{\epsilon \eta} + G^{\alpha \beta \gamma \delta} h_{\gamma \delta} 
    - 8 \pi \bigg[ 2 W^{\alpha \beta \gamma \delta} \nabla_{(\gamma} \xi_{\delta)} -  \nabla_{\gamma} \bigg( T^{\alpha \beta} \xi^{\gamma} \bigg) + 2 T^{\gamma (\beta} \nabla_{\gamma} \xi^{\alpha)} \bigg]
    \nonumber\\
    &- 8 \pi \left(W^{\alpha \beta \gamma \delta} h_{\gamma \delta}\right) \,,\\
    E_{\alpha}[\xi,h] &= 
    (\varepsilon + p) \mathcal{L}_{u} \left( q_{\alpha}^{\beta} u^{\gamma} \Delta g_{\beta \gamma}\right) 
    - 
    \frac{1}{2}\left(\varepsilon +p \right) q_{\alpha}^{\beta} \nabla_{\beta}\left(u^{\gamma}u^{\delta} \Delta g_{\gamma \delta} \right)
    +
    \frac{1}{2} (\nabla_{\alpha} p)\left(1+ \frac{\Gamma p}{\varepsilon + p} \right) q^{\gamma \delta} \Delta g_{\gamma \delta} 
    \nonumber\\
    &- \frac{1}{2} q_{\alpha}^{\beta} \nabla_{\beta}\left( \Gamma p q^{\gamma \delta} \Delta g_{\gamma \delta} \right)
    \,,
\end{align}
\end{subequations}
\end{widetext}
where $\epsilon^{\alpha \beta \gamma \delta}$ is the Levi-Civita tensor and 
\begin{subequations}
\begin{align}
    &G^{\alpha \beta \gamma \delta} \equiv \frac{1}{2} R^{\alpha (\gamma \delta) \beta} +
    \frac{1}{4} R \left(g^{\alpha \gamma} g^{\beta \delta} + g^{\alpha \delta} g^{\beta \gamma} - g^{\alpha \beta} g^{\gamma \delta} \right)
    \nonumber\\
    &
    + 
    \frac{1}{4} \left( 2 R^{\alpha \beta} g^{\gamma \delta} + 2 R^{\gamma \delta} g^{\alpha \beta} - 3 R^{\alpha(\gamma} g^{\delta) \beta}
    - 3 R^{\beta(\gamma} g^{\delta) \alpha}
    \right)
    \,,\\
    &W^{\alpha \beta \gamma \delta} \equiv 
    \frac{1}{2}\left(\varepsilon + p\right) u^{\alpha} u^{\beta} u^{\gamma} u^{\delta}
    \nonumber\\
    &+
    \frac{1}{2} p \left( 
    g^{\alpha \beta} g^{\gamma \delta}
    -
    g^{\alpha\gamma} g^{\beta \delta}
    -
    g^{\alpha \delta} g^{\beta \gamma}
    \right)
    -
    \frac{1}{2}
    \Gamma p q^{\alpha \beta} q^{\gamma \delta}
    \,.
\end{align}
\end{subequations}
The purely ``gravitational'' operators $E^{\alpha \beta}_{\mathrm{grav}}$ and $E^{\alpha}_{\mathrm{grav}}$ used in Eq.~\eqref{eq:tidal-perturbation-v1} of the main text are obtained by setting $\xi_{\mu}= 0$ in Eq.~\eqref{eq:linearized-Einstein-Euler-system}, namely
\begin{subequations}\label{eq:purely-gravitational-operators}
\begin{align}
    &E^{\alpha \beta}_{\mathrm{grav}}[h] \equiv E^{\alpha \beta}[\xi=0,h] = 
    -\frac{1}{2} \epsilon^{\alpha \gamma \epsilon \zeta} \epsilon^{\beta \delta \eta}{}_{\zeta} \nabla_{\gamma} \nabla_{\delta} h_{\epsilon \eta} 
    \nonumber\\
    &\hspace{1cm}
    + 
    G^{\alpha \beta \gamma \delta} h_{\gamma \delta} 
    - 8 \pi \left(W^{\alpha \beta \gamma \delta} h_{\gamma \delta}\right)\,,\\
    &E_{\alpha}^{\mathrm{grav}}[h] \equiv E_{\alpha}[\xi=0,h]
    =
    (\varepsilon + p) \mathcal{L}_{u} \left( q_{\alpha}^{\beta} u^{\gamma} h_{\beta \gamma}\right) 
    \nonumber\\
    &- 
    \frac{1}{2}\left(\varepsilon +p \right) q_{\alpha}^{\beta} \nabla_{\beta}\left(u^{\gamma}u^{\delta} h_{\gamma \delta} \right)
    - \frac{1}{2} q_{\alpha}^{\beta} \nabla_{\beta}\left( \Gamma p q^{\gamma \delta} h_{\gamma \delta} \right)
    \nonumber\\
    &+
    \frac{1}{2} (\nabla_{\alpha} p)\left(1+ \frac{\Gamma p}{\varepsilon + p} \right) q^{\gamma \delta} h_{\gamma \delta} \,.
\end{align}
\end{subequations}
The operator $\mathscr{L}(\hat{y}, y)$ and the function $\Theta^{\beta}$ appearing in Eq.~\eqref{eq:operator-Eqn-FS} of the main text are given by
\begin{subequations}
\begin{align}
    &\mathscr{L}(\hat{y}, y) \equiv 
    U^{\alpha \beta \gamma \delta} \nabla_{\alpha} \hat{\xi}_{\beta} \nabla_{\gamma} \xi_{\delta}
    \nonumber\\
    &+
    V^{\alpha \beta \gamma \delta} \left(\hat{h}_{\alpha \beta} \nabla_{\gamma} \xi_{\delta} + h_{\alpha \beta} \nabla_{\gamma} \hat{\xi}_{\delta}  \right)
    \nonumber\\
    &-
    \frac{1}{32 \pi} \varepsilon^{\alpha \gamma \kappa \sigma} \varepsilon^{\beta \delta \rho}{}_{\sigma} \nabla_{\gamma} \hat{h}_{\alpha \beta} \nabla_{\delta} h_{\kappa \rho}
    -
    T^{\alpha \beta} R_{\alpha \gamma \beta \delta} \hat{\xi}^{\gamma} \xi^{\delta}
    \nonumber\\
    &+
    \left(\frac{1}{2}W^{\alpha \beta \gamma \delta} - \frac{1}{16 \pi} G^{\alpha \beta \gamma \delta} \right) \hat{h}_{\alpha \beta} h_{\gamma \delta}
    \nonumber\\
    &-\frac{1}{2} \nabla_{\gamma} T^{\alpha \beta} \left(\hat{h}_{\alpha \beta} \xi^{\gamma} + h_{\alpha \beta} \hat{\xi}^{\gamma} \right)
    \,,\\
    &\Theta^{\alpha}[\hat{y},y]
    \equiv
    \left[
    U^{\alpha \beta \gamma \delta} \nabla_{\gamma} \xi_{\delta}
    +
    V^{\gamma \delta \alpha \beta} h_{\gamma \delta}
    \right] \hat{\xi}_{\beta} 
    \nonumber\\
    &-\frac{1}{32\pi} \epsilon^{\alpha\epsilon(\gamma}_{\eta} \epsilon^{\beta)\delta \zeta \eta} \hat{h}_{\beta \gamma}\nabla_{\delta} h_{\epsilon \zeta}
    \,,
\end{align}
\end{subequations}
where
\begin{subequations}
\begin{align}
    &U^{\alpha \beta \gamma \delta} \equiv (\varepsilon + p) u^{\alpha} u^{\gamma} q^{\beta \delta} + p (g^{\alpha \beta} g^{\gamma \delta} - g^{\alpha \delta} g^{\beta \gamma}) \nonumber\\
    &
    - \Gamma p q^{\alpha \beta} q^{\gamma \delta} \,,\\
    &2 V^{\alpha \beta \gamma \delta} \equiv
    \left(\varepsilon + p \right)
    \left(
    u^{\alpha} u^{\gamma} q^{\beta \delta}
    +
    u^{\beta} u^{\gamma} q^{\alpha \delta}
    - 
    u^{\alpha} u^{\beta} q^{\gamma \delta}
    \right)
    \nonumber\\
    &-
    \Gamma p q^{\alpha \beta} q^{\gamma \delta}
    \,.
\end{align}
\end{subequations}
The derivations of all the equations provided above can be found in Chapter 7 of~\cite{FN-book}.
\subsection{3+1 split of the operator equation}
Let us present here the derivation of the 3+1 split of the operator equation presented in the main text [Eq.~\eqref{eq:operator-Eqn-FS}].
Without loss of generality, we use the gauge freedom in the definition of the Lagrangian displacement vector $\xi^{\mu}$ to set
\begin{align}
    \xi^{\mu} u_{\mu} = 0\,.
\end{align}
Consider a set of functions $(\mathfrak{h}, \mathfrak{h}_{\rho}, \mathfrak{h}_{\sigma ,\rho})$ that satisfy
\begin{align}
    \mathfrak{h}_{\rho} u^{\rho} = 0 = \mathfrak{h}_{\sigma \rho} u^{\rho} \,,
\end{align}
and define the following operators
\begin{subequations}\label{eq:hamiltonian-and-momentum-constraint-definitions-operator-form}
\begin{align}
    &H \equiv 
    4 \pi \mathfrak{h}^{\alpha }{}_{\alpha } p + 4 \pi \mathfrak{h} p -  \tfrac{1}{2} \mathfrak{h}^{\alpha \beta } \!{}^{(3)}{}R_{\alpha \beta } + \tfrac{1}{4} \mathfrak{h}^{\alpha }{}_{\alpha } \!{}^{(3)}{}R \nonumber \\ 
    & + \tfrac{1}{4} \mathfrak{h} \!{}^{(3)}{}R - 4 \pi \mathfrak{h} h \rho + 8 \pi \xi^{\alpha } \rho D_{\alpha }h - 8 \pi \xi^{\alpha } D_{\alpha }p + 8 \pi h \rho D_{\alpha }\xi^{\alpha } \nonumber \\ 
    & + 8 \pi h \xi^{\alpha } D_{\alpha }\rho + \tfrac{1}{2} D_{\beta }D_{\alpha }\mathfrak{h}^{\alpha \beta } -  \tfrac{1}{2} \gamma^{\alpha \beta } D_{\beta }D_{\alpha }\mathfrak{h}^{\gamma }{}_{\gamma }
    \,,\\
    \label{eq:momentum-constraint-operator}
    &P^{\sigma} \equiv 
    8 \pi \mathfrak{h}^{\sigma } p -  \mathfrak{h}^{\alpha } \!{}^{(3)}{}R^{\sigma }{}_{\alpha } + \tfrac{1}{2} \mathfrak{h}^{\sigma } \!{}^{(3)}{}R + \frac{8 \pi h \xi^{\sigma } \rho}{N} \nonumber \\ 
    & -  \tfrac{1}{2} D_{\alpha }D^{\alpha }\mathfrak{h}^{\sigma } -  \frac{\mathfrak{h}^{\sigma } D_{\alpha }D^{\alpha }N}{2 N} + \tfrac{1}{2} D_{\alpha }D^{\sigma }\mathfrak{h}^{\alpha } \nonumber \\ 
    & + \frac{\mathfrak{h}^{\alpha } D_{\alpha }D^{\sigma }N}{2 N} -  \frac{D_{\alpha }N D^{\alpha }\mathfrak{h}^{\sigma }}{2 N} + \frac{\mathfrak{h}^{\sigma } D_{\alpha }N D^{\alpha }N}{2 N^2} \nonumber \\ 
    & -  \frac{\gamma^{\sigma \beta } D_{\beta }\mathfrak{h}^{\gamma }{}_{\gamma }}{2 N} -  \frac{\mathfrak{h}^{\sigma }{}_{\beta } D^{\beta }N}{2 N^2} + \frac{\gamma^{\sigma \beta } D_{\gamma }\mathfrak{h}_{\beta }{}^{\gamma }}{2 N} \nonumber \\ 
    & + \frac{D_{\alpha }N D^{\sigma }\mathfrak{h}^{\alpha }}{N} + \frac{\mathfrak{h}^{\gamma }{}_{\gamma } D^{\sigma }N}{2 N^2} -  \frac{D_{\alpha }\mathfrak{h}^{\alpha } D^{\sigma }N}{2 N} \nonumber \\ 
    & -  \frac{\mathfrak{h}^{\alpha } D_{\alpha }N D^{\sigma }N}{2 N^2}
    \,.
\end{align}
\end{subequations}
If we split the metric perturbation as
\begin{align}\label{eq:3+1-split-metric}
    h_{\alpha \beta} &\equiv u_{\alpha} u_{\beta} \mathfrak{h} - u_{\alpha} \partial_{t} \mathfrak{h}_{\beta} - u_{\beta} \partial_{t} \mathfrak{h}_{\alpha} + \mathfrak{h}_{\alpha \beta}
\end{align}
where
\begin{subequations}
\begin{align}
    \mathfrak{h} &= h_{\mu \nu} u^{\mu} u^{\nu} \,,\\
    \partial_{t}\mathfrak{h}_{\sigma} &= h_{\mu \nu} u^{\mu} \gamma^{\nu}_{\sigma} \,,\\
    \mathfrak{h}_{\alpha \beta} &= h_{\mu \nu}\gamma^{\nu}_{\beta} \gamma^{\mu}_{\alpha} \,.
\end{align}
\end{subequations}
Then, the operators defined in Eq.~\eqref{eq:hamiltonian-and-momentum-constraint-definitions-operator-form} satisfy the Hamiltonian and momentum constraints of the linearized system
\begin{align*}
    E_{\mu \nu} u^{\mu} u^{\nu} = H\,,  \qquad
    E_{\mu \nu} u^{\mu} \gamma^{\nu}_{\rho} = \partial_{t} P_{\rho} \,.
\end{align*}

To proceed further, we need the 3+1 split of $- \mathscr{L}(\hat{y}, y) + \nabla_{\beta} \Theta^{\beta}[\hat{y}, y]$ and the operator form of $H$ and $P_{\rho}$. 
Let $\hat{y} = (\hat{\xi}, \hat{h}_{\mu\nu})$ be an arbitrary, abstract vector with $\hat{\xi}_{\mu} u^{\mu} = 0$ and 
\begin{align}\label{eq:3+1-split-metric-hat}
    \hat{h}_{\alpha \beta} &\equiv u_{\alpha} u_{\beta} \hat{\mathfrak{h}} - u_{\alpha} \partial_{t} \hat{\mathfrak{h}}_{\beta} - u_{\beta} \partial_{t} \hat{\mathfrak{h}}_{\alpha} + \hat{\mathfrak{h}}_{\alpha \beta}
\end{align}
where
\begin{subequations}
\begin{align}
    \hat{\mathfrak{h}} &= \hat{h}_{\mu \nu} u^{\mu} u^{\nu} \,,\\
    \partial_{t} \hat{\mathfrak{h}}_{\sigma} &= \hat{h}_{\mu \nu} u^{\mu} \gamma^{\nu}_{\sigma} \,,\\
    \hat{\mathfrak{h}}_{\alpha \beta} &= \hat{h}_{\mu \nu}\gamma^{\nu}_{\beta} \gamma^{\mu}_{\alpha} \,.
\end{align}
\end{subequations}
The 3+1 split of $- \mathscr{L}(\hat{y}, y) + \nabla_{\beta} \Theta^{\beta}[\hat{y}, y]$ and the operator form of $H$ and $P_{\rho}$ schematically take the  form
\begin{subequations}
\begin{align}
    \label{eq:operator-form-hamiltonian-general-gauge}
    &\frac{1}{16\pi}\hat{\mathfrak{h}} H = \frac{1}{N}
    \left[\mathcal{A}_{H} + D_{\alpha} \mathcal{R}^{\alpha}_{H}\right] \,,\\
    \label{eq:operator-form-momentum-general-gauge}
    &\frac{1}{8\pi}\hat{\mathfrak{h}}^{\rho} P_{\rho} = 
    \frac{1}{N}
    \left[\mathcal{A}_{P} + D_{\alpha} \mathcal{R}^{\alpha}_{P}\right] \,,\\
    &- \mathscr{L}(\hat{y}, y) + \nabla_{\beta} \Theta^{\beta}[\hat{y}, y]
    =
    \frac{1}{N}
    \left[\mathcal{A}_{L} + D_{\alpha} \mathcal{R}^{\alpha}_{L}\right]
    \,.
\end{align}
\end{subequations}
We can derive these equations by manipulating Eq.~\eqref{eq:hamiltonian-and-momentum-constraint-definitions-operator-form}.
The operators $\mathcal{A}_{H,P,L}$ and $\mathcal{R}^{\alpha}_{H,L,P}$ are given by
\begin{widetext}
\begin{subequations}
\begin{align}
    &\mathcal{A}_{H}[\hat{y},y] = 
    \tfrac{1}{4} \mathfrak{h}^{\alpha }{}_{\alpha } \hat{\mathfrak{h}} p N 
    + \tfrac{1}{4} \mathfrak{h} \hat{\mathfrak{h}} p N 
    -  \frac{\mathfrak{h}^{\alpha \beta } \hat{\mathfrak{h}} \!{}{}{}{}{}^{(3)}{}R_{\alpha \beta } N}{32 \pi} 
    + \frac{\mathfrak{h}^{\alpha }{}_{\alpha } \hat{\mathfrak{h}} \!{}{}{}{}{}^{(3)}{}R N}{64 \pi} 
    + \frac{\mathfrak{h} \hat{\mathfrak{h}} \!{}{}{}{}{}^{(3)}{}R N}{64 \pi} 
    -  \tfrac{1}{4} \mathfrak{h} \hat{\mathfrak{h}} h N \rho 
    + \tfrac{1}{2} \hat{\mathfrak{h}} N \xi^{\alpha } \rho D_{\alpha }h 
    \nonumber\\
    &-  \tfrac{1}{2} \hat{\mathfrak{h}} N \xi^{\alpha } D_{\alpha }p 
    + \tfrac{1}{2} \hat{\mathfrak{h}} h N \rho D_{\alpha }\xi^{\alpha } 
    + \tfrac{1}{2} \hat{\mathfrak{h}} h N \xi^{\alpha } D_{\alpha }\rho 
    + \frac{N D_{\alpha }\mathfrak{h}^{\beta }{}_{\beta } D^{\alpha }\hat{\mathfrak{h}}}{32 \pi} 
    + \frac{\hat{\mathfrak{h}} D_{\alpha }\mathfrak{h}^{\beta }{}_{\beta } D^{\alpha }N}{32 \pi} 
    -  \frac{N D^{\alpha }\hat{\mathfrak{h}} D_{\beta }\mathfrak{h}_{\alpha }{}^{\beta }}{32 \pi} 
    -  \frac{\hat{\mathfrak{h}} D^{\alpha }N D_{\beta }\mathfrak{h}_{\alpha }{}^{\beta }}{32 \pi}\,,\\
    &\mathcal{R}_{H}^{\sigma}[\hat{y},y] = 
    - \frac{\gamma^{\sigma \alpha } \hat{\mathfrak{h}} N D_{\alpha }\mathfrak{h}^{\beta }{}_{\beta }}{32 \pi} + \frac{\hat{\mathfrak{h}} N D_{\alpha }\mathfrak{h}^{\sigma \alpha }}{32 \pi} \,,\\
    &\mathcal{A}_{P}[\hat{y},y]
    =
    \mathfrak{h}^{\alpha } \hat{\mathfrak{h}}_{\alpha } p N -  \frac{3 \mathfrak{h}^{\alpha } \hat{\mathfrak{h}}^{\beta } \!^{(3)}{}R_{\alpha \beta } N}{32 \pi} + \frac{\mathfrak{h}^{\alpha } \hat{\mathfrak{h}}_{\alpha } \!^{(3)}{}R N}{16 \pi} + \hat{\mathfrak{h}}^{\alpha } h \xi_{\alpha } \rho -  \frac{\hat{\mathfrak{h}}^{\alpha } D_{\alpha }\mathfrak{h}^{\beta }{}_{\beta }}{32 \pi} + \frac{\mathfrak{h}^{\beta }{}_{\beta } D_{\alpha }\hat{\mathfrak{h}}^{\alpha }}{32 \pi} \nonumber \\ 
& + \frac{\mathfrak{h}^{\beta }{}_{\beta } \hat{\mathfrak{h}}^{\alpha } D_{\alpha }N}{16 \pi N} + \frac{\hat{\mathfrak{h}}^{\alpha } D_{\beta }\mathfrak{h}_{\alpha }{}^{\beta }}{32 \pi} -  \frac{\hat{\mathfrak{h}}^{\alpha } D_{\alpha }N D_{\beta }\mathfrak{h}^{\beta }}{16 \pi} -  \frac{N D_{\alpha }\mathfrak{h}^{\alpha } D_{\beta }\hat{\mathfrak{h}}^{\beta }}{32 \pi} -  \frac{\mathfrak{h}^{\alpha } D_{\alpha }N D_{\beta }\hat{\mathfrak{h}}^{\beta }}{16 \pi} \nonumber \\ 
& + \frac{\hat{\mathfrak{h}}^{\alpha } D_{\alpha }\mathfrak{h}^{\beta } D_{\beta }N}{32 \pi} + \frac{\mathfrak{h}^{\alpha } D_{\alpha }\hat{\mathfrak{h}}^{\beta } D_{\beta }N}{32 \pi} -  \frac{\mathfrak{h}^{\alpha } \hat{\mathfrak{h}}^{\beta } D_{\alpha }N D_{\beta }N}{16 \pi N} + \frac{\mathfrak{h}^{\alpha } \hat{\mathfrak{h}}^{\beta } D_{\beta }D_{\alpha }N}{32 \pi} -  \frac{\mathfrak{h}^{\alpha } \hat{\mathfrak{h}}_{\alpha } D_{\beta }D^{\beta }N}{32 \pi} \nonumber \\ 
& + \frac{\hat{\mathfrak{h}}^{\alpha } D_{\beta }N D^{\beta }\mathfrak{h}_{\alpha }}{32 \pi} -  \frac{N D_{\alpha }\hat{\mathfrak{h}}_{\beta } D^{\beta }\mathfrak{h}^{\alpha }}{32 \pi} + \frac{N D_{\beta }\hat{\mathfrak{h}}_{\alpha } D^{\beta }\mathfrak{h}^{\alpha }}{16 \pi} + \frac{\mathfrak{h}^{\alpha } D_{\beta }N D^{\beta }\hat{\mathfrak{h}}_{\alpha }}{32 \pi} -  \frac{\mathfrak{h}_{\alpha \beta } D^{\beta }\hat{\mathfrak{h}}^{\alpha }}{32 \pi} -  \frac{\mathfrak{h}_{\alpha \beta } \hat{\mathfrak{h}}^{\alpha } D^{\beta }N}{16 \pi N} \nonumber \\ 
& + \frac{\mathfrak{h}^{\alpha } \hat{\mathfrak{h}}_{\alpha } D_{\beta }N D^{\beta }N}{16 \pi N}
    \,,\\
    &\mathcal{R}_{P}^{\sigma}[\hat{y},y] 
    =
    \frac{\mathfrak{h}^{\sigma }{}_{\alpha } \hat{\mathfrak{h}}^{\alpha }}{32 \pi} -  \frac{\mathfrak{h}^{\alpha }{}_{\alpha } \hat{\mathfrak{h}}^{\sigma }}{32 \pi} + \frac{\hat{\mathfrak{h}}^{\sigma } N D_{\alpha }\mathfrak{h}^{\alpha }}{32 \pi} + \frac{\hat{\mathfrak{h}}^{\alpha } N D_{\alpha }\mathfrak{h}^{\sigma }}{32 \pi} -  \frac{\mathfrak{h}^{\sigma } \hat{\mathfrak{h}}^{\alpha } D_{\alpha }N}{32 \pi} + \frac{\mathfrak{h}^{\alpha } \hat{\mathfrak{h}}^{\sigma } D_{\alpha }N}{16 \pi} \nonumber \\ 
    & -  \frac{\hat{\mathfrak{h}}^{\alpha } N D^{\sigma }\mathfrak{h}_{\alpha }}{16 \pi} -  \frac{\mathfrak{h}^{\alpha } \hat{\mathfrak{h}}_{\alpha } D^{\sigma }N}{32 \pi}
    \,,\\
    &\mathcal{A}_{L}[\hat{y},y]
    =
    \mathcal{A}_{L,0}[\hat{y},y] +\mathcal{A}_{L,1}[\hat{y},y] \,,\nonumber\\
    &\mathcal{A}_{L,0}[\hat{y},y]
    =
- \frac{D^{\alpha }N \partial_t \mathfrak{h}^{\beta }{}_{\beta } \partial_t \hat{\mathfrak{h}}_{\alpha }}{16 \pi N} -  p N \partial_t \mathfrak{h}^{\alpha } \partial_t \hat{\mathfrak{h}}_{\alpha } -  \frac{\!{}{}{}^{(3)}{}R N \partial_t \mathfrak{h}^{\alpha } \partial_t \hat{\mathfrak{h}}_{\alpha }}{16 \pi} + \frac{D_{\beta }D^{\beta }N \partial_t \mathfrak{h}^{\alpha } \partial_t \hat{\mathfrak{h}}_{\alpha }}{32 \pi} + \frac{D^{\alpha }N D^{\beta }N \partial_t \mathfrak{h}_{\alpha } \partial_t \hat{\mathfrak{h}}_{\beta }}{16 \pi N} \nonumber \\ 
& -  \frac{D_{\alpha }N D^{\alpha }N \partial_t \mathfrak{h}^{\beta } \partial_t \hat{\mathfrak{h}}_{\beta }}{16 \pi N} + \frac{D^{\alpha }N \partial_t \mathfrak{h}_{\alpha \beta } \partial_t \hat{\mathfrak{h}}^{\beta }}{16 \pi N} + \frac{3 \!{}{}{}^{(3)}{}R_{\alpha \beta } N \partial_t \mathfrak{h}^{\alpha } \partial_t \hat{\mathfrak{h}}^{\beta }}{32 \pi} -  \frac{D_{\beta }D_{\alpha }N \partial_t \mathfrak{h}^{\alpha } \partial_t \hat{\mathfrak{h}}^{\beta }}{32 \pi} -  h \rho \partial_t \hat{\mathfrak{h}}^{\alpha } \partial_t \xi_{\alpha } \nonumber \\ 
& + \frac{\partial_t \hat{\mathfrak{h}}^{\alpha } \partial_t D_{\alpha }\mathfrak{h}^{\beta }{}_{\beta }}{32 \pi} -  \frac{D^{\alpha }N \partial_t \hat{\mathfrak{h}}^{\beta } \partial_t D_{\alpha }\mathfrak{h}_{\beta }}{32 \pi} -  \frac{\partial_t \mathfrak{h}^{\beta }{}_{\beta } \partial_t D_{\alpha }\hat{\mathfrak{h}}^{\alpha }}{32 \pi} -  \frac{D^{\alpha }N \partial_t \mathfrak{h}^{\beta } \partial_t D_{\alpha }\hat{\mathfrak{h}}_{\beta }}{32 \pi} -  \frac{\partial_t \hat{\mathfrak{h}}^{\alpha } \partial_t D_{\beta }\mathfrak{h}_{\alpha }{}^{\beta }}{32 \pi} \nonumber \\ 
& -  \frac{D^{\alpha }N \partial_t \hat{\mathfrak{h}}^{\beta } \partial_t D_{\beta }\mathfrak{h}_{\alpha }}{32 \pi} + \frac{D^{\alpha }N \partial_t \hat{\mathfrak{h}}_{\alpha } \partial_t D_{\beta }\mathfrak{h}^{\beta }}{16 \pi} -  \frac{D^{\alpha }N \partial_t \mathfrak{h}^{\beta } \partial_t D_{\beta }\hat{\mathfrak{h}}_{\alpha }}{32 \pi} + \frac{D^{\alpha }N \partial_t \mathfrak{h}_{\alpha } \partial_t D_{\beta }\hat{\mathfrak{h}}^{\beta }}{16 \pi} + \frac{N \partial_t D_{\alpha }\mathfrak{h}^{\alpha } \partial_t D_{\beta }\hat{\mathfrak{h}}^{\beta }}{32 \pi} \nonumber \\ 
& + \frac{N \partial_t D_{\alpha }\hat{\mathfrak{h}}_{\beta } \partial_t D^{\beta }\mathfrak{h}^{\alpha }}{32 \pi} -  \frac{N \partial_t D_{\beta }\hat{\mathfrak{h}}_{\alpha } \partial_t D^{\beta }\mathfrak{h}^{\alpha }}{16 \pi} + \frac{\partial_t \mathfrak{h}_{\alpha \beta } \partial_t D^{\beta }\hat{\mathfrak{h}}^{\alpha }}{32 \pi} + \frac{\hat{\mathfrak{h}}^{\alpha \beta } \partial_t^{2} \mathfrak{h}_{\alpha \beta }}{32 \pi N} -  \frac{\hat{\mathfrak{h}}^{\alpha }{}_{\alpha } \partial_t^{2} \mathfrak{h}^{\beta }{}_{\beta }}{32 \pi N} \nonumber \\ 
& + h \hat{\xi}^{\alpha } \rho \partial_t^{2} \mathfrak{h}_{\alpha } + \frac{\hat{\mathfrak{h}}^{\beta }{}_{\beta } D^{\alpha }N \partial_t^{2} \mathfrak{h}_{\alpha }}{16 \pi N} -  \frac{D_{\alpha }\hat{\mathfrak{h}}^{\beta }{}_{\beta } \partial_t^{2} \mathfrak{h}^{\alpha }}{32 \pi} + \frac{D_{\beta }\hat{\mathfrak{h}}_{\alpha }{}^{\beta } \partial_t^{2} \mathfrak{h}^{\alpha }}{32 \pi} -  \frac{\hat{\mathfrak{h}}_{\alpha \beta } D^{\alpha }N \partial_t^{2} \mathfrak{h}^{\beta }}{16 \pi N} + \frac{h \hat{\xi}^{\alpha } \rho \partial_t^{2} \xi_{\alpha }}{N} \nonumber \\ 
& + \frac{\hat{\mathfrak{h}}^{\beta }{}_{\beta } \partial_t^{2} D_{\alpha }\mathfrak{h}^{\alpha }}{32 \pi} -  \frac{\hat{\mathfrak{h}}_{\alpha \beta } \partial_t^{2} D^{\beta }\mathfrak{h}^{\alpha }}{32 \pi}
\,,\nonumber\\
&\mathcal{A}_{L,1}[\hat{y},y]
=
\tfrac{1}{2} \mathfrak{h}^{\alpha \beta } \hat{\mathfrak{h}}_{\alpha \beta } p N -  \tfrac{1}{4} \mathfrak{h}^{\alpha }{}_{\alpha } \hat{\mathfrak{h}}^{\beta }{}_{\beta } p N + \tfrac{1}{4} \hat{\mathfrak{h}}^{\alpha }{}_{\alpha } \mathfrak{h} p N + \tfrac{1}{4} \mathfrak{h}^{\alpha }{}_{\alpha } \hat{\mathfrak{h}} p N + \tfrac{1}{4} \mathfrak{h} \hat{\mathfrak{h}} p N + \frac{\mathfrak{h}^{\alpha \beta } \hat{\mathfrak{h}}^{\gamma }{}_{\gamma } \!{}{}^{(3)}{}R_{\alpha \beta } N}{32 \pi} \nonumber \\ 
& -  \frac{\hat{\mathfrak{h}}^{\alpha \beta } \mathfrak{h} \!{}{}^{(3)}{}R_{\alpha \beta } N}{32 \pi} -  \frac{\mathfrak{h}^{\alpha \beta } \hat{\mathfrak{h}} \!{}{}^{(3)}{}R_{\alpha \beta } N}{32 \pi} -  \frac{3 \mathfrak{h}^{\alpha \beta } \hat{\mathfrak{h}}_{\alpha }{}^{\gamma } \!{}{}^{(3)}{}R_{\beta \gamma } N}{32 \pi} + \frac{\mathfrak{h}^{\alpha }{}_{\alpha } \hat{\mathfrak{h}}^{\beta \gamma } \!{}{}^{(3)}{}R_{\beta \gamma } N}{32 \pi} + \frac{\mathfrak{h}^{\alpha \beta } \hat{\mathfrak{h}}_{\alpha \beta } \!{}{}^{(3)}{}R N}{32 \pi} \nonumber \\ 
& -  \frac{\mathfrak{h}^{\alpha }{}_{\alpha } \hat{\mathfrak{h}}^{\beta }{}_{\beta } \!{}{}^{(3)}{}R N}{64 \pi} + \frac{\hat{\mathfrak{h}}^{\alpha }{}_{\alpha } \mathfrak{h} \!{}{}^{(3)}{}R N}{64 \pi} + \frac{\mathfrak{h}^{\alpha }{}_{\alpha } \hat{\mathfrak{h}} \!{}{}^{(3)}{}R N}{64 \pi} + \frac{\mathfrak{h} \hat{\mathfrak{h}} \!{}{}^{(3)}{}R N}{64 \pi} -  \frac{\mathfrak{h}^{\alpha \beta } \hat{\mathfrak{h}}^{\gamma \delta }{}{}^{(3)}{}R_{\alpha \gamma \beta \delta } N}{32 \pi} \nonumber \\ 
& + \tfrac{1}{4} \mathfrak{h}^{\alpha }{}_{\alpha } \hat{\mathfrak{h}}^{\beta }{}_{\beta } p N \Gamma + p \!{}{}^{(3)}{}R_{\alpha \beta } N \xi^{\alpha } \hat{\xi}^{\beta } -  \tfrac{1}{4} \mathfrak{h} \hat{\mathfrak{h}} h N \rho + \tfrac{1}{2} \hat{\mathfrak{h}} N \xi^{\alpha } \rho D_{\alpha }h + \tfrac{1}{2} \mathfrak{h} N \hat{\xi}^{\alpha } \rho D_{\alpha }h + \tfrac{1}{2} \hat{\mathfrak{h}}^{\beta }{}_{\beta } N \xi^{\alpha } D_{\alpha }p -  \tfrac{1}{2} \hat{\mathfrak{h}} N \xi^{\alpha } D_{\alpha }p \nonumber \\ 
& + \tfrac{1}{2} \mathfrak{h}^{\beta }{}_{\beta } N \hat{\xi}^{\alpha } D_{\alpha }p -  \tfrac{1}{2} \mathfrak{h} N \hat{\xi}^{\alpha } D_{\alpha }p + \tfrac{1}{2} \hat{\mathfrak{h}}^{\beta }{}_{\beta } p N \Gamma D_{\alpha }\xi^{\alpha } + \tfrac{1}{2} \hat{\mathfrak{h}} h N \rho D_{\alpha }\xi^{\alpha } + \tfrac{1}{2} \mathfrak{h}^{\beta }{}_{\beta } p N \Gamma D_{\alpha }\hat{\xi}^{\alpha } + \tfrac{1}{2} \mathfrak{h} h N \rho D_{\alpha }\hat{\xi}^{\alpha } + \tfrac{1}{2} \hat{\mathfrak{h}} h N \xi^{\alpha } D_{\alpha }\rho \nonumber \\ 
& + \tfrac{1}{2} \mathfrak{h} h N \hat{\xi}^{\alpha } D_{\alpha }\rho + \frac{N D_{\alpha }\hat{\mathfrak{h}}^{\beta }{}_{\beta } D^{\alpha }\mathfrak{h}}{32 \pi} + \frac{N D_{\alpha }\mathfrak{h}^{\beta }{}_{\beta } D^{\alpha }\hat{\mathfrak{h}}}{32 \pi} + \frac{\hat{\mathfrak{h}} D_{\alpha }\mathfrak{h}^{\beta }{}_{\beta } D^{\alpha }N}{32 \pi} + \frac{\mathfrak{h} D_{\alpha }\hat{\mathfrak{h}}^{\beta }{}_{\beta } D^{\alpha }N}{32 \pi} -  \frac{N D^{\alpha }\hat{\mathfrak{h}} D_{\beta }\mathfrak{h}_{\alpha }{}^{\beta }}{32 \pi} \nonumber \\ 
& -  \frac{\hat{\mathfrak{h}} D^{\alpha }N D_{\beta }\mathfrak{h}_{\alpha }{}^{\beta }}{32 \pi} + \frac{\hat{\mathfrak{h}}_{\alpha }{}^{\beta } D^{\alpha }N D_{\beta }\mathfrak{h}^{\gamma }{}_{\gamma }}{32 \pi} -  \frac{N D^{\alpha }\mathfrak{h} D_{\beta }\hat{\mathfrak{h}}_{\alpha }{}^{\beta }}{32 \pi} -  \frac{\mathfrak{h} D^{\alpha }N D_{\beta }\hat{\mathfrak{h}}_{\alpha }{}^{\beta }}{32 \pi} +
\frac{\gamma^{\alpha \beta } N D_{\alpha }\mathfrak{h}^{\gamma \delta } D_{\beta }\hat{\mathfrak{h}}_{\gamma \delta }}{32 \pi}  + \frac{N D_{\alpha }\mathfrak{h}^{\alpha \beta } D_{\beta }\hat{\mathfrak{h}}^{\gamma }{}_{\gamma }}{32 \pi} 
\nonumber \\ 
&+ \frac{\mathfrak{h}_{\alpha }{}^{\beta } D^{\alpha }N D_{\beta }\hat{\mathfrak{h}}^{\gamma }{}_{\gamma }}{32 \pi} 
-  \frac{\gamma^{\alpha \beta } N D_{\alpha }\mathfrak{h}^{\gamma }{}_{\gamma } D_{\beta }\hat{\mathfrak{h}}^{\delta }{}_{\delta }}{32 \pi} -  p \hat{\xi}^{\alpha } D_{\alpha }N D_{\beta }\xi^{\beta } 
-  p \xi^{\alpha } D_{\alpha }N D_{\beta }\hat{\xi}^{\beta } \nonumber \\ 
& -  p N D_{\alpha }\xi^{\alpha } D_{\beta }\hat{\xi}^{\beta } + p N \Gamma D_{\alpha }\xi^{\alpha } D_{\beta }\hat{\xi}^{\beta } -  \frac{\mathfrak{h}^{\alpha \beta } \hat{\mathfrak{h}}^{\gamma }{}_{\gamma } D_{\beta }D_{\alpha }N}{32 \pi} -  p \xi^{\alpha } \hat{\xi}^{\beta } D_{\beta }D_{\alpha }N + h \xi^{\alpha } \hat{\xi}^{\beta } \rho D_{\beta }D_{\alpha }N + p N D_{\alpha }\hat{\xi}_{\beta } D^{\beta }\xi^{\alpha } 
\nonumber \\ 
&-  \frac{\hat{\mathfrak{h}}_{\alpha }{}^{\beta } D^{\alpha }N D_{\gamma }\mathfrak{h}_{\beta }{}^{\gamma }}{32 \pi}  
-  \frac{N D_{\alpha }\mathfrak{h}^{\alpha \beta } D_{\gamma }\hat{\mathfrak{h}}_{\beta }{}^{\gamma }}{32 \pi} -  \frac{\mathfrak{h}_{\alpha }{}^{\beta } D^{\alpha }N D_{\gamma }\hat{\mathfrak{h}}_{\beta }{}^{\gamma }}{32 \pi} + \frac{N D^{\beta }\mathfrak{h}^{\alpha }{}_{\alpha } D_{\gamma }\hat{\mathfrak{h}}_{\beta }{}^{\gamma }}{32 \pi} + \frac{3 \mathfrak{h}^{\alpha \beta } \hat{\mathfrak{h}}_{\alpha }{}^{\gamma } D_{\gamma }D_{\beta }N}{32 \pi} \nonumber \\ 
& -  \frac{\mathfrak{h}^{\alpha }{}_{\alpha } \hat{\mathfrak{h}}^{\beta \gamma } D_{\gamma }D_{\beta }N}{32 \pi} -  \frac{\mathfrak{h}^{\alpha \beta } \hat{\mathfrak{h}}_{\alpha \beta } D_{\gamma }D^{\gamma }N}{16 \pi} + \frac{\mathfrak{h}^{\alpha }{}_{\alpha } \hat{\mathfrak{h}}^{\beta }{}_{\beta } D_{\gamma }D^{\gamma }N}{32 \pi} -  \frac{N D_{\beta }\hat{\mathfrak{h}}_{\alpha \gamma } D^{\gamma }\mathfrak{h}^{\alpha \beta }}{32 \pi}
\,,\\
&\mathcal{R}_{L}^{\sigma}[\hat{y},y]
=
\mathcal{R}_{L,0}^{\sigma}[\hat{y},y] +\mathcal{R}_{L,1}^{\sigma}[\hat{y},y] \,,\nonumber\\
&\mathcal{R}_{L,0}^{\sigma}[\hat{y},y] =
\frac{D^{\sigma }N \partial_t \mathfrak{h}^{\alpha } \partial_t \hat{\mathfrak{h}}_{\alpha }}{32 \pi} + \frac{D^{\alpha }N \partial_t \mathfrak{h}^{\sigma } \partial_t \hat{\mathfrak{h}}_{\alpha }}{32 \pi} -  \frac{\partial_t \mathfrak{h}^{\sigma }{}_{\alpha } \partial_t \hat{\mathfrak{h}}^{\alpha }}{32 \pi} + \frac{\partial_t \mathfrak{h}^{\alpha }{}_{\alpha } \partial_t \hat{\mathfrak{h}}^{\sigma }}{32 \pi} -  \frac{D^{\alpha }N \partial_t \mathfrak{h}_{\alpha } \partial_t \hat{\mathfrak{h}}^{\sigma }}{16 \pi} \nonumber \\ 
& -  \frac{N \partial_t \hat{\mathfrak{h}}^{\sigma } \partial_t D_{\alpha }\mathfrak{h}^{\alpha }}{32 \pi} -  \frac{N \partial_t \hat{\mathfrak{h}}^{\alpha } \partial_t D_{\alpha }\mathfrak{h}^{\sigma }}{32 \pi} + \frac{N \partial_t \hat{\mathfrak{h}}^{\alpha } \partial_t D^{\sigma }\mathfrak{h}_{\alpha }}{16 \pi} -  \frac{\hat{\mathfrak{h}}^{\sigma }{}_{\alpha } \partial_t^{2} \mathfrak{h}^{\alpha }}{32 \pi} + \frac{\hat{\mathfrak{h}}^{\alpha }{}_{\alpha } \partial_t^{2} \mathfrak{h}^{\sigma }}{32 \pi}
\,,\nonumber\\
&\mathcal{R}_{L,1}^{\sigma}[\hat{y},y] =
- \tfrac{1}{2} \mathfrak{h}^{\alpha }{}_{\alpha } p N \Gamma \hat{\xi}^{\sigma } -  \tfrac{1}{2} \mathfrak{h} h N \hat{\xi}^{\sigma } \rho -  \frac{\gamma^{\sigma \alpha } \hat{\mathfrak{h}}^{\beta \gamma } N D_{\alpha }\mathfrak{h}_{\beta \gamma }}{32 \pi} -  \frac{\hat{\mathfrak{h}}^{\sigma \alpha } N D_{\alpha }\mathfrak{h}^{\beta }{}_{\beta }}{32 \pi} -  \frac{\gamma^{\sigma \alpha } \hat{\mathfrak{h}} N D_{\alpha }\mathfrak{h}^{\beta }{}_{\beta }}{32 \pi} + \frac{\gamma^{\sigma \alpha } \hat{\mathfrak{h}}^{\beta }{}_{\beta } N D_{\alpha }\mathfrak{h}^{\gamma }{}_{\gamma }}{32 \pi} \nonumber \\ 
& + \frac{\hat{\mathfrak{h}} N D_{\alpha }\mathfrak{h}^{\sigma \alpha }}{32 \pi} + p \xi^{\alpha } \hat{\xi}^{\sigma } D_{\alpha }N + p N \hat{\xi}^{\sigma } D_{\alpha }\xi^{\alpha } -  p N \Gamma \hat{\xi}^{\sigma } D_{\alpha }\xi^{\alpha } -  p N \hat{\xi}^{\alpha } D_{\alpha }\xi^{\sigma } + \frac{\hat{\mathfrak{h}}^{\sigma }{}_{\alpha } N D^{\alpha }\mathfrak{h}}{32 \pi} -  \frac{\mathfrak{h}^{\sigma }{}_{\alpha } \hat{\mathfrak{h}}^{\beta }{}_{\beta } D^{\alpha }N}{32 \pi} + \frac{\mathfrak{h}_{\alpha }{}^{\beta } \hat{\mathfrak{h}}^{\sigma }{}_{\beta } D^{\alpha }N}{32 \pi} 
\nonumber \\ 
& + \frac{\hat{\mathfrak{h}}^{\sigma }{}_{\alpha } \mathfrak{h} D^{\alpha }N}{32 \pi} + \frac{\hat{\mathfrak{h}}^{\sigma \alpha } N D_{\beta }\mathfrak{h}_{\alpha }{}^{\beta }}{32 \pi} + \frac{\hat{\mathfrak{h}}^{\alpha \beta } N D_{\beta }\mathfrak{h}^{\sigma }{}_{\alpha }}{32 \pi} -  \frac{\hat{\mathfrak{h}}^{\alpha }{}_{\alpha } N D_{\beta }\mathfrak{h}^{\sigma \beta }}{32 \pi} -  \frac{\hat{\mathfrak{h}}^{\alpha }{}_{\alpha } N D^{\sigma }\mathfrak{h}}{32 \pi}
-  
\frac{\hat{\mathfrak{h}}^{\alpha }{}_{\alpha } \mathfrak{h} D^{\sigma }N}{32 \pi}
\,.
\end{align}
\end{subequations}
\end{widetext}

With these identities in hand, we can derive the 3+1 split of the general operator equation provided in Eq.~\eqref{eq:operator-Eqn-FS} of the main text by using that the Hamiltonian and momentum constraints are satisfied. 
Suppose $\hat{\mathfrak{h}}$, $\hat{\mathfrak{h}}_{\rho}$, and $\hat{\mathfrak{h}}_{\rho,\sigma}$ satisfy the Hamiltonian and the momentum constraints $\hat{H} = 0 = \hat{P}_{\rho}$.
Consider two abstract vectors $\hat{y} = (\hat{\xi}_{\mu},\hat{h}_{\mu\nu})$ and $y = (\xi_{\mu},h_{\mu\nu})$, where the metric perturbations are split in 3+1 form, as given in Eqs.~\eqref{eq:3+1-split-metric} and \eqref{eq:3+1-split-metric-hat}.
Suppose the abstract vector $y$ is a solution to the linearized Einstein-Euler system. Then, by using Eq.~\eqref{eq:operator-Eqn-FS} of the main text, we conclude that the following equation is satisfied
\begin{align}\label{eq:E-operator-supplementary}
    \mathcal{E}[\hat{y}, y]
    \equiv &N\bigg[-\mathscr{L}(\hat{y}, y) + \nabla_{\beta} \Theta^{\beta}[\hat{y}, y] +
    \frac{\partial_t \hat{\mathfrak{h}}_{\rho}}{8\pi}
    \partial_t P^{\rho}
    \nonumber\\
    &-
    \frac{1}{8\pi}
    \hat{P}^{\rho}
    \partial_t^2 \mathfrak{h}^{\rho}
    \bigg]
    =\mathcal{A} + D_{\alpha}\mathcal{R}^{\alpha}  = 0\,.
\end{align}
The operator $\mathcal{A}$ has the following schematic form
\begin{align}\label{eq:A-operator-general-gauge-supplementary}
    \mathcal{A} &= \hat{y}_{A} O_{0}^{AB} \partial_t^2 y_{B} + \hat{y}_{A} O_{1}^{AB} y_{B} \,,   
\end{align}
where $A$ is a general index used to label the component of the abstract vector $y$.
The operator $O_{0}^{AB}$ and $O_{1}^{AB}$ are symmetric operators
\begin{widetext}
\begin{subequations}\label{eq:operator-general-case}
\begin{align}
    &\hat{y}_{A} O_{0}^{AB} y_{B} \equiv 
    \frac{\mathfrak{h}^{\alpha \beta } \hat{\mathfrak{h}}_{\alpha \beta }}{32 \pi N} -  \frac{\mathfrak{h}^{\alpha }{}_{\alpha } \hat{\mathfrak{h}}^{\beta }{}_{\beta }}{32 \pi N} -  \mathfrak{h}^{\alpha } \hat{\mathfrak{h}}_{\alpha } p N + \frac{3 \mathfrak{h}^{\alpha } \hat{\mathfrak{h}}^{\beta } \,{}{}{}{}{}^{(3)}R_{\alpha \beta } N}{32 \pi} -  \frac{\mathfrak{h}^{\alpha } \hat{\mathfrak{h}}_{\alpha } \,{}{}{}{}{}^{(3)}R N}{16 \pi} + \frac{h \xi^{\alpha } \hat{\xi}_{\alpha } \rho}{N} + \frac{\hat{\mathfrak{h}}^{\alpha } D_{\alpha }N D_{\beta }\mathfrak{h}^{\beta }}{16 \pi} + \frac{N D_{\alpha }\mathfrak{h}^{\alpha } D_{\beta }\hat{\mathfrak{h}}^{\beta }}{32 \pi} 
    \nonumber\\
    &+ \frac{\mathfrak{h}^{\alpha } D_{\alpha }N D_{\beta }\hat{\mathfrak{h}}^{\beta }}{16 \pi} -  \frac{\hat{\mathfrak{h}}^{\alpha } D_{\alpha }\mathfrak{h}^{\beta } D_{\beta }N}{32 \pi} -  \frac{\mathfrak{h}^{\alpha } D_{\alpha }\hat{\mathfrak{h}}^{\beta } D_{\beta }N}{32 \pi} + \frac{\mathfrak{h}^{\alpha } \hat{\mathfrak{h}}^{\beta } D_{\alpha }N D_{\beta }N}{16 \pi N} -  \frac{\mathfrak{h}^{\alpha } \hat{\mathfrak{h}}^{\beta } D_{\beta }D_{\alpha }N}{32 \pi} + \frac{\mathfrak{h}^{\alpha } \hat{\mathfrak{h}}_{\alpha } D_{\beta }D^{\beta }N}{32 \pi} 
    \nonumber\\
    &-  \frac{\hat{\mathfrak{h}}^{\alpha } D_{\beta }N D^{\beta }\mathfrak{h}_{\alpha }}{32 \pi} + \frac{N D_{\alpha }\hat{\mathfrak{h}}_{\beta } D^{\beta }\mathfrak{h}^{\alpha }}{32 \pi} -  \frac{N D_{\beta }\hat{\mathfrak{h}}_{\alpha } D^{\beta }\mathfrak{h}^{\alpha }}{16 \pi} 
    -  \frac{\mathfrak{h}^{\alpha } D_{\beta }N D^{\beta }\hat{\mathfrak{h}}_{\alpha }}{32 \pi} -  \frac{\mathfrak{h}^{\alpha } \hat{\mathfrak{h}}_{\alpha } D_{\beta }N D^{\beta }N}{16 \pi N}
    \,,\\
    &\hat{y}_{A} O_{1}^{AB} y_{B} \equiv 
    \tfrac{1}{2} \mathfrak{h}^{\alpha \beta } \hat{\mathfrak{h}}_{\alpha \beta } p N -  \tfrac{1}{4} \mathfrak{h}^{\alpha }{}_{\alpha } \hat{\mathfrak{h}}^{\beta }{}_{\beta } p N + \tfrac{1}{4} \hat{\mathfrak{h}}^{\alpha }{}_{\alpha } \mathfrak{h} p N + \tfrac{1}{4} \mathfrak{h}^{\alpha }{}_{\alpha } \hat{\mathfrak{h}} p N + \tfrac{1}{4} \mathfrak{h} \hat{\mathfrak{h}} p N + \frac{\mathfrak{h}^{\alpha \beta } \hat{\mathfrak{h}}^{\gamma }{}_{\gamma } \!{}^{(3)}{}R_{\alpha \beta } N}{32 \pi} \nonumber \\ 
    & -  \frac{\hat{\mathfrak{h}}^{\alpha \beta } \mathfrak{h} \!{}^{(3)}{}R_{\alpha \beta } N}{32 \pi} -  \frac{\mathfrak{h}^{\alpha \beta } \hat{\mathfrak{h}} \!{}^{(3)}{}R_{\alpha \beta } N}{32 \pi} -  \frac{3 \mathfrak{h}^{\alpha \beta } \hat{\mathfrak{h}}_{\alpha }{}^{\gamma } \!{}^{(3)}{}R_{\beta \gamma } N}{32 \pi} + \frac{\mathfrak{h}^{\alpha }{}_{\alpha } \hat{\mathfrak{h}}^{\beta \gamma } \!{}^{(3)}{}R_{\beta \gamma } N}{32 \pi} + \frac{\mathfrak{h}^{\alpha \beta } \hat{\mathfrak{h}}_{\alpha \beta } \!{}^{(3)}{}R N}{32 \pi} \nonumber \\ 
    & -  \frac{\mathfrak{h}^{\alpha }{}_{\alpha } \hat{\mathfrak{h}}^{\beta }{}_{\beta } \!{}^{(3)}{}R N}{64 \pi} + \frac{\hat{\mathfrak{h}}^{\alpha }{}_{\alpha } \mathfrak{h} \!{}^{(3)}{}R N}{64 \pi} + \frac{\mathfrak{h}^{\alpha }{}_{\alpha } \hat{\mathfrak{h}} \!{}^{(3)}{}R N}{64 \pi} + \frac{\mathfrak{h} \hat{\mathfrak{h}} \!{}^{(3)}{}R N}{64 \pi} -  \frac{\mathfrak{h}^{\alpha \beta } \hat{\mathfrak{h}}^{\gamma \delta }{}^{(3)}{}R_{\alpha \gamma \beta \delta } N}{32 \pi} \nonumber \\ 
    & + \tfrac{1}{4} \mathfrak{h}^{\alpha }{}_{\alpha } \hat{\mathfrak{h}}^{\beta }{}_{\beta } p N \Gamma + p \!{}^{(3)}{}R_{\alpha \beta } N \xi^{\alpha } \hat{\xi}^{\beta } -  \tfrac{1}{4} \mathfrak{h} \hat{\mathfrak{h}} h N \rho + \tfrac{1}{2} \hat{\mathfrak{h}} N \xi^{\alpha } \rho D_{\alpha }h + \tfrac{1}{2} \mathfrak{h} N \hat{\xi}^{\alpha } \rho D_{\alpha }h + \tfrac{1}{2} \hat{\mathfrak{h}}^{\beta }{}_{\beta } N \xi^{\alpha } D_{\alpha }p -  \tfrac{1}{2} \hat{\mathfrak{h}} N \xi^{\alpha } D_{\alpha }p \nonumber \\ 
    & + \tfrac{1}{2} \mathfrak{h}^{\beta }{}_{\beta } N \hat{\xi}^{\alpha } D_{\alpha }p -  \tfrac{1}{2} \mathfrak{h} N \hat{\xi}^{\alpha } D_{\alpha }p + \tfrac{1}{2} \hat{\mathfrak{h}}^{\beta }{}_{\beta } p N \Gamma D_{\alpha }\xi^{\alpha } + \tfrac{1}{2} \hat{\mathfrak{h}} h N \rho D_{\alpha }\xi^{\alpha } + \tfrac{1}{2} \mathfrak{h}^{\beta }{}_{\beta } p N \Gamma D_{\alpha }\hat{\xi}^{\alpha } + \tfrac{1}{2} \mathfrak{h} h N \rho D_{\alpha }\hat{\xi}^{\alpha } + \tfrac{1}{2} \hat{\mathfrak{h}} h N \xi^{\alpha } D_{\alpha }\rho \nonumber \\ 
    & + \tfrac{1}{2} \mathfrak{h} h N \hat{\xi}^{\alpha } D_{\alpha }\rho + \frac{N D_{\alpha }\hat{\mathfrak{h}}^{\beta }{}_{\beta } D^{\alpha }\mathfrak{h}}{32 \pi} + \frac{N D_{\alpha }\mathfrak{h}^{\beta }{}_{\beta } D^{\alpha }\hat{\mathfrak{h}}}{32 \pi} + \frac{\hat{\mathfrak{h}} D_{\alpha }\mathfrak{h}^{\beta }{}_{\beta } D^{\alpha }N}{32 \pi} + \frac{\mathfrak{h} D_{\alpha }\hat{\mathfrak{h}}^{\beta }{}_{\beta } D^{\alpha }N}{32 \pi} -  \frac{N D^{\alpha }\hat{\mathfrak{h}} D_{\beta }\mathfrak{h}_{\alpha }{}^{\beta }}{32 \pi} \nonumber \\ 
    & -  \frac{\hat{\mathfrak{h}} D^{\alpha }N D_{\beta }\mathfrak{h}_{\alpha }{}^{\beta }}{32 \pi} + \frac{\hat{\mathfrak{h}}_{\alpha }{}^{\beta } D^{\alpha }N D_{\beta }\mathfrak{h}^{\gamma }{}_{\gamma }}{32 \pi} -  \frac{N D^{\alpha }\mathfrak{h} D_{\beta }\hat{\mathfrak{h}}_{\alpha }{}^{\beta }}{32 \pi} -  \frac{\mathfrak{h} D^{\alpha }N D_{\beta }\hat{\mathfrak{h}}_{\alpha }{}^{\beta }}{32 \pi} + \frac{N D_{\alpha }\mathfrak{h}^{\alpha \beta } D_{\beta }\hat{\mathfrak{h}}^{\gamma }{}_{\gamma }}{32 \pi} \nonumber \\ 
    & + \frac{\mathfrak{h}_{\alpha }{}^{\beta } D^{\alpha }N D_{\beta }\hat{\mathfrak{h}}^{\gamma }{}_{\gamma }}{32 \pi} -  p \hat{\xi}^{\alpha } D_{\alpha }N D_{\beta }\xi^{\beta } -  p \xi^{\alpha } D_{\alpha }N D_{\beta }\hat{\xi}^{\beta } -  p N D_{\alpha }\xi^{\alpha } D_{\beta }\hat{\xi}^{\beta } + p N \Gamma D_{\alpha }\xi^{\alpha } D_{\beta }\hat{\xi}^{\beta } -  \frac{\mathfrak{h}^{\alpha \beta } \hat{\mathfrak{h}}^{\gamma }{}_{\gamma } D_{\beta }D_{\alpha }N}{32 \pi} \nonumber \\ 
    & -  p \xi^{\alpha } \hat{\xi}^{\beta } D_{\beta }D_{\alpha }N + h \xi^{\alpha } \hat{\xi}^{\beta } \rho D_{\beta }D_{\alpha }N -  \frac{N D_{\beta }\hat{\mathfrak{h}}^{\gamma }{}_{\gamma } D^{\beta }\mathfrak{h}^{\alpha }{}_{\alpha }}{32 \pi} + p N D_{\alpha }\hat{\xi}_{\beta } D^{\beta }\xi^{\alpha } -  \frac{\hat{\mathfrak{h}}_{\alpha }{}^{\beta } D^{\alpha }N D_{\gamma }\mathfrak{h}_{\beta }{}^{\gamma }}{32 \pi} -  \frac{N D_{\alpha }\mathfrak{h}^{\alpha \beta } D_{\gamma }\hat{\mathfrak{h}}_{\beta }{}^{\gamma }}{32 \pi} \nonumber \\ 
    & -  \frac{\mathfrak{h}_{\alpha }{}^{\beta } D^{\alpha }N D_{\gamma }\hat{\mathfrak{h}}_{\beta }{}^{\gamma }}{32 \pi} + \frac{N D^{\beta }\mathfrak{h}^{\alpha }{}_{\alpha } D_{\gamma }\hat{\mathfrak{h}}_{\beta }{}^{\gamma }}{32 \pi} + \frac{3 \mathfrak{h}^{\alpha \beta } \hat{\mathfrak{h}}_{\alpha }{}^{\gamma } D_{\gamma }D_{\beta }N}{32 \pi} -  \frac{\mathfrak{h}^{\alpha }{}_{\alpha } \hat{\mathfrak{h}}^{\beta \gamma } D_{\gamma }D_{\beta }N}{32 \pi} \nonumber \\ 
    & -  \frac{\mathfrak{h}^{\alpha \beta } \hat{\mathfrak{h}}_{\alpha \beta } D_{\gamma }D^{\gamma }N}{16 \pi} + \frac{\mathfrak{h}^{\alpha }{}_{\alpha } \hat{\mathfrak{h}}^{\beta }{}_{\beta } D_{\gamma }D^{\gamma }N}{32 \pi} -  \frac{N D_{\beta }\hat{\mathfrak{h}}_{\alpha \gamma } D^{\gamma }\mathfrak{h}^{\alpha \beta }}{32 \pi} + \frac{N D_{\gamma }\hat{\mathfrak{h}}_{\alpha \beta } D^{\gamma }\mathfrak{h}^{\alpha \beta }}{32 \pi}
    \,.
\end{align}
\end{subequations}
\end{widetext}
The symmetry properties of the operator $\mathcal{R}^{\alpha}$ on the other hand are not obvious in a general gauge
\begin{widetext}
\begin{align}
    \mathcal{R}^{\sigma}
    &=
    - \tfrac{1}{2} \mathfrak{h}^{\alpha }{}_{\alpha } p N \Gamma\hat{\xi}^{\sigma } -  \tfrac{1}{2} \mathfrak{h} h N \hat{\xi}^{\sigma } \rho -  \frac{\hat{\mathfrak{h}}^{\sigma \alpha } N D_{\alpha }\mathfrak{h}^{\beta }{}_{\beta }}{32 \pi} + \frac{\hat{\mathfrak{h}} N D_{\alpha }\mathfrak{h}^{\sigma \alpha }}{32 \pi} + p \xi^{\alpha } \hat{\xi}^{\sigma } D_{\alpha }N + p N \hat{\xi}^{\sigma } D_{\alpha }\xi^{\alpha } -  p N \Gamma\hat{\xi}^{\sigma } D_{\alpha }\xi^{\alpha } 
    \nonumber\\
    &-  p N \hat{\xi}^{\alpha } D_{\alpha }\xi^{\sigma } 
    + \frac{\hat{\mathfrak{h}}^{\sigma }{}_{\alpha } N D^{\alpha }\mathfrak{h}}{32 \pi} -  \frac{\mathfrak{h}^{\sigma }{}_{\alpha } \hat{\mathfrak{h}}^{\beta }{}_{\beta } D^{\alpha }N}{32 \pi} + \frac{\mathfrak{h}_{\alpha }{}^{\beta } \hat{\mathfrak{h}}^{\sigma }{}_{\beta } D^{\alpha }N}{32 \pi} 
    \nonumber\\
    &+ \frac{\hat{\mathfrak{h}}^{\sigma }{}_{\alpha } \mathfrak{h} D^{\alpha }N}{32 \pi} 
    + \frac{\hat{\mathfrak{h}}^{\sigma \alpha } N D_{\beta }\mathfrak{h}_{\alpha }{}^{\beta }}{32 \pi} + \frac{\hat{\mathfrak{h}}^{\alpha \beta } N D_{\beta }\mathfrak{h}^{\sigma }{}_{\alpha }}{32 \pi} -  \frac{\hat{\mathfrak{h}}^{\alpha }{}_{\alpha } N D_{\beta }\mathfrak{h}^{\sigma \beta }}{32 \pi} -  \frac{\hat{\mathfrak{h}}^{\alpha \beta } N D^{\sigma }\mathfrak{h}_{\alpha \beta }}{32 \pi} -  \frac{\hat{\mathfrak{h}} N D^{\sigma }\mathfrak{h}^{\alpha }{}_{\alpha }}{32 \pi} 
    \nonumber\\
    &+ \frac{\hat{\mathfrak{h}}^{\alpha }{}_{\alpha } N D^{\sigma }\mathfrak{h}^{\beta }{}_{\beta }}{32 \pi} -  \frac{\hat{\mathfrak{h}}^{\alpha }{}_{\alpha } N D^{\sigma }\mathfrak{h}}{32 \pi} -  \frac{\hat{\mathfrak{h}}^{\alpha }{}_{\alpha } \mathfrak{h} D^{\sigma }N}{32 \pi} -  \frac{N D^{\alpha }\hat{\mathfrak{h}}^{\sigma } \partial_t^{2} \mathfrak{h}_{\alpha }}{32 \pi} + \frac{\hat{\mathfrak{h}}^{\sigma } D^{\alpha }N \partial_t^{2} \mathfrak{h}_{\alpha }}{32 \pi} + \frac{N D^{\sigma }\hat{\mathfrak{h}}^{\alpha } \partial_t^{2} \mathfrak{h}_{\alpha }}{16 \pi} 
    \nonumber\\
    &+ \frac{\hat{\mathfrak{h}}^{\alpha } D^{\sigma }N \partial_t^{2} \mathfrak{h}_{\alpha }}{32 \pi} -  \frac{\hat{\mathfrak{h}}^{\sigma }{}_{\alpha } \partial_t^{2} \mathfrak{h}^{\alpha }}{16 \pi} 
    + \frac{\hat{\mathfrak{h}}^{\alpha }{}_{\alpha } \partial_t^{2} \mathfrak{h}^{\sigma }}{16 \pi} -  \frac{N D_{\alpha }\hat{\mathfrak{h}}^{\alpha } \partial_t^{2} \mathfrak{h}^{\sigma }}{32 \pi} -  \frac{\hat{\mathfrak{h}}^{\alpha } D_{\alpha }N \partial_t^{2} \mathfrak{h}^{\sigma }}{16 \pi}
    \,.
\end{align}
\end{widetext}
\subsection{Proof in harmonic gauge}\label{sec:harmonic-gauge}
We now study the operator equation [Eq.~\eqref{eq:E-operator-supplementary}] in the harmonic gauge.
Define the scalar fields 
\begin{subequations}\label{eq:background-harmonic-coordinate}
\begin{align}
    X^{(0)} &= t\,,\\
    X^{(1)} &= R \sin(\theta) \cos(\phi) \,,\\
    X^{(2)} &= R \sin(\theta) \sin(\phi) \,,\\
    X{}{}{}{}^{(3)} &= R \cos(\theta) \,,
\end{align}
\end{subequations}
where $(t,r,\theta,\phi)$ are the usual Schwarzschild coordinates, and $R=r-M$.
These scalar fields are harmonic functions in the Schwarzschild spacetime exterior to the star 
\begin{align}\label{eq:harmonic-condition-background}
    \Box_{g}{X^{(\mu)}} = 0 \,.
\end{align}
Let us work with $(t, R, \theta,\phi) $ coordinates outside the star.
We also demand that the scalar fields $X^{(\mu)}$ be harmonic functions in the perturbed spacetime exterior to the star,
\begin{align}\label{eq:harmonic-gauge-conditions}
    \Box_{g+h}{X^{(\mu)}} = 0\implies \nabla_{\alpha} \left[\bar{h}^{\alpha \beta} \nabla_{\beta} X^{(\mu)} \right] = 0\,,
\end{align}
where the trace reversed metric perturbation is given by 
\begin{align}\label{eq:harmonic-gauge-condition}
    \Bar{h}_{\mu\nu} = h_{\mu\nu} - \frac{1}{2} h^{\alpha}_{\alpha} \, g_{\mu \nu}\,.
\end{align}
Observe that Eq.~\eqref{eq:harmonic-gauge-conditions} is a covariant form of the familiar harmonic gauge conditions imposed in PN theory~\cite{Blanchet:2013haa,Taylor_2008}.

As we described in the main text, we demand that the solutions to the linearized equations match asymptotically in the buffer-zone to the PN solutions, which are valid in the PN zone.
The buffer zone is assumed to be located at a distance $R=d$ (see Fig.~\ref{fig:zones} in the main text).
Let $Y_{\ell m}$, $E_{A,\ell m}$ and $Z_{AB,\ell m}$ denote the scalar, polar vector and polar tensor spherical harmonics, and $\Omega_{AB}$ denote the metric on the 2-sphere\footnote{In this section, we use capital Latin letters to denote the components of tensors along the $(\theta,\phi)$ directions of spacetime. These indices should \textit{not} be confused with the components of the abstract vector $y$, used in the main body of this paper.}. 
The PN solutions in the buffer zone accurate to 1PN order for a particular $(\ell,m)$ spherical mode with multipole moment $I_{\ell m}(t)$ are 
\begin{subequations}\label{eq:PN-boundary-condition-harmonic-gauge}
\begin{align}
    &\bar{h}_{tt} = \frac{16\pi}{(2\ell+1) c^2} \bigg[\frac{I_{\ell m} }{R^{\ell+1}} 
    \nonumber\\
    &\hspace{1cm}
    - 
    \frac{1}{c^2}\bigg( 
    \frac{5 I_{\ell m} M}{2 R^{\ell+2}}
    + 
    \frac{\partial_{t}^2 I_{\ell m} }{2 (2 \ell-1) R^{\ell-1}}
    \bigg)\bigg] Y_{\ell m}\,,\\
    &\bar{h}_{t R} = \frac{16 \pi  \partial_{t} I_{\ell m} }{(1-2 \ell) (2 \ell+1) R^{\ell} c^2} Y_{\ell m} \,, \\
    &\bar{h}_{t A} = \frac{16 \pi \partial_{t}I_{\ell m} R^{1-\ell}}{(2 \ell+1) \left(\ell-2 \ell^2\right) c^2} E_{A,\ell m}\,, \\
    &\bar{h}_{RR} = \frac{8 \pi  I_{\ell m} M }{(2 \ell+1)R^{\ell+2}c^4} Y_{\ell m}  
    + \frac{16 \pi Y_{\ell m} R^{-\ell+1} \partial_{t}^2I_{\ell m} }{(8 \ell^3-12 \ell^2-2 \ell+3) c^4}  \,,\\
    &\bar{h}_{R A} = -\frac{1}{c^4}\bigg[\frac{8 \pi  I_{\ell m} M R^{-\ell-1}}{2 \ell^2+3 \ell+1}
    \nonumber\\
    &\hspace{1cm}
    -\frac{16 \pi R^{2-\ell} \partial_{t}^2I_{\ell m} }{\ell \left(8 \ell^3-12 \ell^2-2 \ell+3\right)}\bigg] E_{A,\ell m}\,, \\
    &\bar{h}_{AB} = \frac{Z_{AB,\ell m}}{c^4} \bigg[ 
     \frac{16\pi M I_{\ell m} R^{-\ell}}{(\ell^2+3 \ell+2)(2 \ell +1)}
     \nonumber\\
     &\hspace{1cm}
     -
     \frac{16 \pi R^{-\ell+3}  \partial_{t}^2I_{\ell m}}{(-4 \ell^4+12 \ell^3-11 \ell^2+3 \ell)(2 \ell+1)}
    \bigg] \nonumber\\
    &\hspace{1cm}
    -
    \frac{\Omega_{AB}Y_{\ell m}}{c^4}
    \Bigg[
    \frac{8 \pi R^{3-\ell} \partial_{t}^2I_{\ell m}  }{(2 \ell-3) (2 \ell-1) (2 \ell+1)}
    \Bigg]
    \,.
\end{align}
\end{subequations}
We have introduced factors of $c$ in the above equations to denote the PN order of the expressions. 
These solutions can be obtained from~\cite{Racine:2004xg}, where general solutions, including non-linear tidal interactions, are provided.
[see also App.~B of~\cite{PhysRevD.46.4304} and Box 7.5 of~\cite{Poisson-Will}].
There are tidal contributions arising from the external environment that we have ignored in the above equations. We provide these solutions when we understand forced oscillations arising from tidally induced perturbations in Sec.~\ref{sec:forced-perturbation}. 

We now show that the mode solutions to the linearized Einstein-Euler system arise from a self-adjoint operator.
Define a new operator $\mathcal{E}_{\mathrm{harm}}$
\begin{align}
    &\mathcal{E}_{\mathrm{harm}} \equiv 
    \mathcal{E} 
    -\frac{N}{32 \pi} \left[ \hat{\mathfrak{h}} H + \mathfrak{h} \hat{H} \right]
    \nonumber\\
    &+\frac{N(10 \ell^2+3 \ell-2)}{16 \pi \left(4 \ell^2+2 \ell-1\right)} \bigg[ 
    \hat{\mathfrak{h}}^{\sigma}\partial_{t}^2 P_{\sigma}  
    + 
    \hat{P}_{\sigma}\partial_{t}^2 \mathfrak{h}^{\sigma} 
    \bigg] = 0\,,
\end{align}
where $\ell$ is the spherical harmonic mode.
One can show that 
\begin{align}\label{eq:E-operator-harm}
    \mathcal{E}_{\mathrm{harm}} &= \mathcal{A}_{\mathrm{harm}} + D_{\sigma} \mathcal{R}_{\mathrm{harm}}^{\sigma}\,.
\end{align}
The operators $\mathcal{A}_{\mathrm{harm}}$ is of the form
\begin{align}
    \mathcal{A}_{\mathrm{harm}} &= \hat{y}_{A} O_{0,\mathrm{harm}}^{AB} \partial_t^2 y_{B} + \hat{y}_{A} O_{1,\mathrm{harm}}^{AB} y_{B}
\end{align}
where the operators $\hat{y}_{A} O_{0,\mathrm{harm}}^{AB} y_{B}$ and $\hat{y}_{A} O_{1,\mathrm{harm}}^{AB}y_{B}$ are symmetric operators 
\begin{subequations}
\begin{align}
    &\hat{y}_{A} O_{0,\mathrm{harm}}^{AB} y_{B} = \hat{y}_{A} O_{0}^{AB} y_{B} 
    \nonumber\\
    &+\frac{(10 \ell^2+3 \ell-2)}{4 \left(4 \ell^2+2 \ell-1\right)} \left[ \mathcal{A}_{P}[\hat{y},y] + \mathcal{A}_{P}[y,\hat{y}] \right]\,,\\
    &\hat{y}_{A} O_{1,\mathrm{harm}}^{AB} y_{B} = \hat{y}_{A} O_{1}^{AB} y_{B} -\frac{1}{2} \left[ \mathcal{A}_{H}[\hat{y},y] + \mathcal{A}_{H}[y,\hat{y}] \right]
\end{align}
\end{subequations}
and
\begin{widetext}
\begin{align}\label{eq:R-harm-expr}
    &\mathcal{R}_{\mathrm{harm}}^{\sigma}[\hat{y},y]
    =
    \mathcal{R}^{\sigma}[\hat{y},y]
    +
    \bigg[
    \frac{\gamma^{\sigma \alpha } \hat{\mathfrak{h}} N D_{\alpha }\mathfrak{h}^{\beta }{}_{\beta }}{64 \pi} -  \frac{\hat{\mathfrak{h}} N D_{\alpha }\mathfrak{h}^{\sigma \alpha }}{64 \pi} + \frac{\gamma^{\sigma \alpha } \mathfrak{h} N D_{\alpha }\hat{\mathfrak{h}}^{\beta }{}_{\beta }}{64 \pi} -  \frac{\mathfrak{h} N D_{\alpha }\hat{\mathfrak{h}}^{\sigma \alpha }}{64 \pi}
    \bigg]
    \nonumber\\
    &+
    \frac{10 \ell^2+3 \ell-2}{2 \left(4 \ell^2+2 \ell-1\right)}
    \bigg[ 
    - \frac{\hat{\mathfrak{h}}^{\sigma } \partial_t^{2} \mathfrak{h}^{\alpha }{}_{\alpha }}{64 \pi} + \frac{\hat{\mathfrak{h}}^{\alpha } \partial_t^{2} \mathfrak{h}^{\sigma }{}_{\alpha }}{64 \pi} + \frac{N D^{\alpha }\hat{\mathfrak{h}}^{\sigma } \partial_t^{2} \mathfrak{h}_{\alpha }}{64 \pi} + \frac{\hat{\mathfrak{h}}^{\sigma } D^{\alpha }N \partial_t^{2} \mathfrak{h}_{\alpha }}{64 \pi} -  \frac{N D^{\sigma }\hat{\mathfrak{h}}^{\alpha } \partial_t^{2} \mathfrak{h}_{\alpha }}{32 \pi} -  \frac{\hat{\mathfrak{h}}^{\alpha } D^{\sigma }N \partial_t^{2} \mathfrak{h}_{\alpha }}{32 \pi} \nonumber \\ 
    &+ \frac{\hat{\mathfrak{h}}^{\sigma }{}_{\alpha } \partial_t^{2} \mathfrak{h}^{\alpha }}{64 \pi} -  \frac{\hat{\mathfrak{h}}^{\alpha }{}_{\alpha } \partial_t^{2} \mathfrak{h}^{\sigma }}{64 \pi} + \frac{N D_{\alpha }\hat{\mathfrak{h}}^{\alpha } \partial_t^{2} \mathfrak{h}^{\sigma }}{64 \pi} + \frac{\hat{\mathfrak{h}}^{\alpha } D_{\alpha }N \partial_t^{2} \mathfrak{h}^{\sigma }}{64 \pi} + \frac{\hat{\mathfrak{h}}^{\sigma } N \partial_t^{2} D_{\alpha }\mathfrak{h}^{\alpha }}{64 \pi} + \frac{\hat{\mathfrak{h}}^{\alpha } N \partial_t^{2} D_{\alpha }\mathfrak{h}^{\sigma }}{64 \pi} \nonumber \\ 
    &-  \frac{\hat{\mathfrak{h}}^{\alpha } N \partial_t^{2} D^{\sigma }\mathfrak{h}_{\alpha }}{32 \pi}
        \bigg]
\,.
\end{align}
\end{widetext}
To obtain a symmetric eigenvalue problem, we work in the frequency domain and integrate Eq.~\eqref{eq:E-operator-harm} over a spatial hypersurface that extends from the center of the star to the mean radius of the buffer zone $R= d$,
\begin{align}\label{eq:harmonic-operator-v1}
    &\int \mathcal{E}_{\mathrm{harm}} \sqrt{\gamma} d^3 x 
    =
    \oint_{R=d} \mathcal{R}^{\sigma}_{\mathrm{harm}} n_{\sigma}(R + M)^2 d \Omega
    \nonumber\\
    &
    -\int \bigg[\omega^2 \hat{y}_{A} O_{0,\mathrm{harm}}^{AB} y_{B} - \hat{y}_{A} O_{1,\mathrm{harm}}^{AB} y_{B} \bigg]\sqrt{\gamma} d^3 x
    \,,
\end{align}
where the normal covector to constant-$R$ hypersurfaces is given by 
\begin{align}
    n_{\sigma} = \left(0, \left({1-\frac{2M}{R+M}}\right)^{-1/2},0,0\right)\,.
\end{align}
The operator on the second line of Eq.~\eqref{eq:harmonic-operator-v1} is symmetric because of the symmetry properties of $O_{1,\mathrm{harm}}^{AB}$ and $O_{0,\mathrm{harm}}^{AB}$.
Using Eq.~\eqref{eq:PN-boundary-condition-harmonic-gauge} and \eqref{eq:R-harm-expr}, we can show that 
\begin{align}\label{eq:boundary-term-harmonic}
    \oint_{R=d} \mathcal{R}^{\sigma}_{\mathrm{harm}}  n_{\sigma}(R + M)^2 d \Omega 
    &=
    \frac{2 \pi  I_{\ell m} \hat{I}_{\ell m} (\ell-2) M}{c^2 (2 \ell+1)^2 d^{2 (\ell+1)}} \nonumber\\
    &+ \mathcal{O}(c^{-4}) \,.
\end{align}
Combining Eqs.~\eqref{eq:harmonic-operator-v1} and \eqref{eq:boundary-term-harmonic}, we see that mode solutions satisfy the following operator equation:
\begin{align}\label{eq:mode-eqn-harmonic}
    &\omega^2 \int \sqrt{\gamma} d^3 x\left[ \hat{y}_{A} O_{0,\mathrm{harm}}^{AB} y_{B}\right]
    =
    \int \sqrt{\gamma} d^3 x \left[\hat{y}_{A} O_{1,\mathrm{harm}}^{AB} y_{B}\right]
    \nonumber\\
    &\hspace{1cm}
    +
    \left[\frac{2 \pi  I_{\ell m} \hat{I}_{\ell m} (\ell-2) M}{c^2 (2 \ell+1)^2 d^{2 (\ell+1)}}+ \mathcal{O}(c^{-4}) \right]
\end{align}
This equation holds true for any general vector $\hat{y}$ that satisfies the Hamiltonian and momentum constraints.
Since both operators on the left- and right-hand sides are symmetric operators, we conclude that the system is self-adjoint.
\subsection{Comments on the proof and comparison to Newtonian theory}
Let us now compare our proof to the techniques used in Newtonian and PN theory~\cite{1964ApJ...139..664C,1965ApJ...142.1519C,gittins2025perturbationtheorypostnewtonianneutron}.
In Newtonian theory, the perturbations to the gravitational potential $U_{\mathrm{Newt}}$ are completely governed by the Poisson equation,
\begin{align}
    \nabla^2 U_{\mathrm{Newt}} = - 4 \pi \delta \rho_{\mathrm{Newt}}\,,
\end{align}
and one can obtain an explicit solution for $U_{\mathrm{New}}$ using Green's function,
\begin{align}\label{eq:U-newt-sol}
    U_{\mathrm{Newt}} &= \int \frac{\delta \rho_{\mathrm{Newt}}}{|x-x'|} d^3 x' \,.
\end{align}
The presence of this explicit solution allows one to analyze the linearized gravitational and fluid perturbations in terms of the Lagrangian displacement vector alone. Moreover, the analysis can be restricted to the interior of the fluid star, since the solution in Eq.~\eqref{eq:U-newt-sol} is valid both inside the star and satisfies the correct boundary condition outside the star. 
In PN theory, one can similarly obtain explicit expressions for the PN potentials using Green's functions~\cite{1965ApJ...142.1519C,gittins2025perturbationtheorypostnewtonianneutron,yin2025postnewtonianapproachneutronstar}, and obtain the symmetry properties of the operator governing the perturbed equations.
However, in GR, one cannot obtain explicit analytical solutions inside the star, and, therefore, we are forced to work with the solution outside the star and carefully treat the boundary conditions. 
Moreover, the gravitational field is also dynamical in GR and one cannot analyze the perturbed equations only in terms of the Lagrangian displacement vector.

We also note that our work is not based on global PN expansions i.e., in our work, the PN approximation is only valid in the buffer-zone (see, Fig.~\ref{fig:zones} in the main text) far away from the star. 
The physics in the inner and the outer body zone is treated without making any PN approximations unlike~\cite{gittins2025perturbationtheorypostnewtonianneutron,yin2025postnewtonianapproachneutronstar}, where a PN approximation is also used in the inner and outer body zone.

From Eq.~\eqref{eq:mode-eqn-harmonic} and Eq.~\eqref{eq:normalization-condition-star} of the main text, it might appear the modes are sensitive to the precise location of the buffer zone. 
This is true at the formal level, but at a practical level, one can use the analytical solutions for the gravitational perturbations exterior to the star~\cite{Poisson:2020vap,HegadeKR:2024agt} to construct an operator equation where the domain of integration is restricted only to the interior of the star. We present the details of this practical implementation in~\cite{future-work-in-prep}.

Finally, we note that our work relies on an implicit assumption that the operator $O_{0,\mathrm{harm}}$ is invertible. In Newtonian gravity, one can indeed show that this is the case. However, there is no formal proof of this assumption when PN corrections are present. One can (and we did) check that the operator is invertible numerically~\cite{1965ApJ...142.1519C,gittins2025perturbationtheorypostnewtonianneutron}. Therefore, we assume that the operator $O_{0,\mathrm{harm}}$ is invertible in this work, and we leave the numerical demonstration of this fact to future work~\cite{future-work-in-prep}.
\subsection{Validity of the 1PN boundary conditions and extension to higher PN order}
In this section, we briefly discuss the errors introduced in the description of tidal interactions due to the use of PN boundary conditions and outline the steps required to extend the boundary conditions to higher PN orders.

To quantify the error in our approximation, it is important to note that we have employed a matched asymptotic expansion, with the asymptotic boundary condition imposed at a mean radius $d$ in the buffer zone. This naively implies that the errors are sensitive to the matching radius $d$. 
In reality, however, the matching radius is only used to ensure that the solution to the linearized Einstein equations in the outer body zone matches consistently to a PN solution~\cite{Poisson:2020vap,HegadeKR:2024agt}.
Once this matching is done, we can use the full relativistic near-zone solution to impose the boundary condition at the \textit{surface of the star}.
Such near-zone boundary conditions have been shown to lead to an estimate of the $f$-mode quasi-normal mode frequencies of neutron stars that is accurate to about $0.1\%$ (see Fig. 4 of~\cite{Lindblom_1997}) and understand the dynamical tidal response of neutron stars, see Figs. 2-3 of~\cite{HegadeKR:2024agt}.

The details of constructing the external solution and numerical integration of the linearized field equations to obtain the modes and the dynamical tidal response will be presented elsewhere~\cite{future-work-in-prep}.
To estimate the error in our approximation, we note that the near-zone solution is obtained using a small-frequency expansion of the Regge-Wheeler equation and is accurate to $\mathcal{O}( \omega^2)$.
Therefore, the error in our approximation depends on the frequency $\omega$ and the radius $R$ of the star and scales approximately as $(R \omega)^4$.
We expect these frequency-dependent corrections to the near-zone solution to show up as corrections to the relativistic factor $\mathcal{Q}(\ell,z_0,\omega)$, the operators $O_{0,1}$, and the mode frequencies $\omega_s$ in Eq.~(20).
We can use this scaling relationship of the approximation to estimate the error. 
For simplicity, assume that we are studying the dynamical tidal response of a neutron star of mass $2 M_{\odot}$ with radius $\sim 10\, \mathrm{km}$.
In the early inspiral, the star could experience $g$-mode excitations with $ R\omega\sim 0.06$; hence, the error is approximately $\sim \mathcal{O}(10^{-5})$ during a $g$-mode resonance.
During the late inspiral, the frequency increases, and we can use $R\omega \sim 0.25$. The error in this case is approximately equal to $\sim (0.25)^4 \sim \mathcal{O}(10^{-3})$. 

Finally, let us comment on the technical steps needed to extend the matched asymptotic expansion to higher PN orders. 
The use of matched asymptotic expansion to determine the tidal response requires four main ingredients: i) a solution to the linearized Einstein equation in a small-frequency approximation, ii) a PN metric in the buffer zone which includes tidal interactions, iii) consistent matching between the PN metric and the near-zone solution and iv) construction of the operator equation to determine the mode frequencies.
In this paper, we have used a 1PN metric in the buffer zone to construct the operator equation and determine the operator equation.
We will present the details of the construction of the near-zone solution to $\mathcal{O}(\omega^2)$, and we will consistently match this solution to the PN metric to determine the modes in a future publication~\cite{future-work-in-prep}.
Extending the near-zone boundary condition to higher orders in frequency will require a solution to the Regge-Wheeler equation up to $\omega^4$, the development of a PN theory of tidal interactions accurate to 2PN order, and a consistent matching between the near-zone solution and the buffer zone solutions in harmonic gauge. 
Once these technical calculations are completed, we expect that the formalism outlined in this paper can be used with the higher-order PN boundary conditions to obtain the tidal overlap integral and the mode frequencies, as the dynamics are conservative in the near-zone.
\section{Tidal perturbations and the dynamical tidal response function}\label{sec:forced-perturbation}
\subsection{Expression for the tidal contributions to the metric perturbation in a PN expansion }
The tidal part of the PN solution accurate to 1PN order in $(t,R,\theta,\phi)$ coordinates is given by
\begin{subequations}\label{eq:PN-boundary-conditions-tide-harmonic-gauge}
\begin{align}
    &\bar{h}_{tt}^{\mathrm{T,ext}} = 
    \frac{16 \pi  d_{\ell m} R^{\ell} Y_{\ell m}}{(2 {\ell}+1)c^2} 
    +
    \frac{Y_{\ell m}}{c^4}
    \bigg(
    -\frac{40 \pi  d_{\ell m} M R^{{\ell}-1}}{(2 {\ell}+1)}
    \nonumber\\
    &\hspace{1cm}+
    \frac{8 \pi  \partial_t^2d_{\ell m} R^{{\ell}+2}}{ \left(4 {\ell}^2+8 {\ell}+3\right)}
    \bigg) \,,\\
    &\bar{h}_{tR}^{\mathrm{T,ext}} = 
    \frac{16 \pi  \partial_{t}d_{\ell m} R^{{\ell}+1} Y_{\ell m}}{(4 {\ell}^2+8 {\ell}+3)c^2}
    \,,\\
    &\bar{h}_{tA}^{\mathrm{T,ext}} = 
    -\frac{16 \pi  \partial_{t}d_{\ell m} R^{{\ell}+2}}{(4 {\ell}^3+12 {\ell}^2+11 {\ell}+3)c^2} E_{A,\ell m}\,,\\
    &\bar{h}_{RR}^{\mathrm{T,ext}} =
    \frac{8 \pi  d_{\ell m} M R^{{\ell}-1} Y_{\ell m}}{(2 {\ell}+1)c^4} 
    +
    \frac{16 \pi Y_{\ell m} \partial_{t}^2d_{\ell m}  R^{{\ell}+2}}{(8 {\ell}^3+36 {\ell}^2+46 {\ell}+15)c^4} \,,\\
    &\bar{h}_{RA}^{\mathrm{T,ext}} = 
    \frac{8 \pi  d_{\ell m} M R^{\ell} E_{A,\ell m}}{(2 {\ell}^2+{\ell})c^4}
    \nonumber\\
    &\hspace{1cm}
    -
    \frac{16 \pi  E_{A,\ell m} \partial_{t}^2d_{\ell m}  R^{{\ell}+3}}{(8 {\ell}^4+44 {\ell}^3+82 {\ell}^2+61 {\ell}+15)c^4}
    \,,\\
    &\bar{h}_{AB}^{\mathrm{T,ext}} = 
    -\Omega_{AB} Y_{\ell m}
    \left[ 
    \frac{8 \pi  \partial_{t}^2d_{\ell m}  R^{{\ell}+4}}{c^4 (2 {\ell}+1) (2 {\ell}+3) (2 {\ell}+5)}
    \right]
    \nonumber\\
    &+
    \frac{Z_{AB,\ell m}}{c^4}
    \bigg[ 
    -\frac{16 \pi  d_{\ell m} M R^{{\ell}+1}}{-2 {\ell}^3+{\ell}^2+{\ell}}
    \nonumber\\
    &\hspace{1cm}
    +
    \frac{16 \pi  \partial_{t}^2d_{\ell m} R^{{\ell}+4}}{8 {\ell}^5+60 {\ell}^4+170 {\ell}^3+225 {\ell}^2+137 {\ell}+30}
    \bigg]
    \,,
\end{align}
\end{subequations}
where $d_{\ell m}(t)$ are the tidal moments. Expressions for the tidal moments in terms of the symmetric trace-free tensors can be found in Eq.~(2.265) of~\cite{Poisson-Will}.
\subsection{Derivation of the formula for the dynamical tidal response function}
Let us outline the steps required to obtain Eq.~\eqref{eq:dynamical-tidal-response-func-GR} of the main text. 
The Hamiltonian constraint for the tidal field is given by
\begin{align}
    &E_{\alpha \beta}^{\mathrm{grav}}u^{\mu} u^{\nu}
    =
    H[h^{T}] =
    4 \pi \mathfrak{h}^{\mathrm{T }\alpha }{}_{\alpha } p + 4 \pi \mathfrak{h}^{\mathrm{T}} p -  \tfrac{1}{2} \mathfrak{h}^{\mathrm{T } \alpha \beta } \,{}{}{}{}{}^{(3)}R_{\alpha \beta } 
    \nonumber\\
    &+ \tfrac{1}{4} \mathfrak{h}^{\mathrm{T } \alpha }{}_{\alpha } \,{}{}{}{}{}^{(3)}R 
    + \tfrac{1}{4} \mathfrak{h}^{\mathrm{T}} \,{}{}{}{}{}^{(3)}R 
    - 4 \pi \mathfrak{h}^{\mathrm{T}} h \rho  
    + \tfrac{1}{2} D_{\beta }D_{\alpha }\mathfrak{h}^{\mathrm{T }\alpha \beta } 
    \nonumber\\
    &-  \tfrac{1}{2} \gamma^{\alpha \beta } D_{\beta }D_{\alpha }\mathfrak{h}^{\mathrm{T }\gamma }{}_{\gamma }
    \,.
\end{align}
Using the above equation and the definition of the Hamiltonian constraint for $h_{\mu\nu}$ [Eq.~\eqref{eq:hamiltonian-and-momentum-constraint-definitions-operator-form}], we see that the following identity is valid
\begin{align}\label{eq:hamiltonian-constraint-operator-Love}
    0 
    =
    \frac{1}{32 \pi}
    \left( H[h^{T}]\bar{\mathfrak{h}}-H[h] \bar{\mathfrak{h}}^{T}\right)
    =
    \mathcal{A}_{\mathrm{Love}} 
    +
    D_{\alpha} \mathcal{R}^{\alpha}_{\mathrm{Love}}
    \,.
\end{align}
The operators $\mathcal{A}_{\mathrm{Love}} $ and $\mathcal{R}^{\alpha}_{\mathrm{Love}}$ are given by
\begin{subequations}\label{eq:A-love-R-love-operators}
\begin{align}
    \mathcal{A}_{\mathrm{Love}} 
    &=
    \tfrac{1}{4} h \xi^{\alpha } \rho D_{\alpha }\bar{\mathfrak{h}}^{T}{}
    +
    \tfrac{1}{4} \bar{\mathfrak{h}}^{T}{} \xi^{\alpha } D_{\alpha }p
    \nonumber\\
    &+
    \frac{\!{}{}{}{}{}^{(3)}{}G^{\alpha \beta } \bar{\mathfrak{h}}_{\alpha \beta } \bar{\mathfrak{h}}^{T}{}}{64 \pi}
    - \frac{\!{}{}{}{}{}^{(3)}{}G^{\alpha \beta } \bar{\mathfrak{h}}^{T}{}_{\alpha \beta } \bar{\mathfrak{h}}}{64 \pi} 
    \nonumber\\
    &+
    \tfrac{1}{16} \bar{\mathfrak{h}}^{\alpha }{}_{\alpha } \bar{\mathfrak{h}}^{T}{} h \rho 
    - \tfrac{1}{16} \bar{\mathfrak{h}}^{T}{}^{\alpha }{}_{\alpha } \bar{\mathfrak{h}} h \rho 
    \nonumber\\
    &+
    \frac{D^{\alpha }\bar{\mathfrak{h}}^{T}{} D_{\beta }\bar{\mathfrak{h}}_{\alpha }{}^{\beta }}{64 \pi} -  \frac{D^{\alpha }\bar{\mathfrak{h}} D_{\beta }\bar{\mathfrak{h}}^{T}{}_{\alpha }{}^{\beta }}{64 \pi}
    \,,\\
    \mathcal{R}_{\alpha}^{\mathrm{Love}}
    &=
    - \tfrac{1}{4} \bar{\mathfrak{h}}^{T}{} h \xi_{\alpha } \rho + \frac{\bar{\mathfrak{h}}^{T}{} D_{\alpha }\bar{\mathfrak{h}}}{64 \pi} -  \frac{\bar{\mathfrak{h}} D_{\alpha }\bar{\mathfrak{h}}^{T}{}}{64 \pi} 
    \nonumber\\
    &-  \frac{\bar{\mathfrak{h}}^{T}{} D_{\sigma }\bar{\mathfrak{h}}_{\alpha }{}^{\sigma }}{64 \pi} + \frac{\bar{\mathfrak{h}} D_{\sigma }\bar{\mathfrak{h}}^{T}{}_{\alpha }{}^{\sigma }}{64 \pi}
    \,.
\end{align}
\end{subequations}
Equation~\eqref{eq:hamiltonian-constraint-operator-Love} is valid for any abstract vector $y^A = (\xi, h_{\mu \nu})$.
Let us then set $y^A = (\xi^{*}, h_{\mu\nu}^{*})$, transform to the Fourier domain, and integrate Eq.~\eqref{eq:hamiltonian-constraint-operator-Love} over the domain of the star; we then obtain
\begin{align}
    \int_{\Sigma_{\mathrm{star}}} \sqrt{\gamma} d^3x \mathcal{A}_{\mathrm{Love}}
    &=
    -\oint_{R=R_{\star}} \mathcal{R}^{\alpha}_{\mathrm{Love}} n_{\alpha} (R+M)^2 d \Omega\,. 
\end{align}
This equation can be simplified by assuming a harmonic time dependence and using Eq.~\eqref{eq:A-love-R-love-operators} and Eqs.~\eqref{eq:Is-def} and \eqref{eq:sol-a-freq} of the main text. 
The left-hand side of this equation simplifies to
\begin{widetext}
\begin{align}\label{eq:LHS-for-love}
    \int_{\Sigma_{\mathrm{star}}} \sqrt{\gamma} d^3x \mathcal{A}_{\mathrm{Love}}
    &=
    \sum_{s}  \tilde{a}_{s}(\omega)
    \int_{\Sigma_{\mathrm{star}}} \sqrt{\gamma} d^3 x \bigg[ 
    \tfrac{1}{4} h \xi^{*\alpha }_s \rho D_{\alpha }\bar{\Tilde{\mathfrak{h}}}^{T}{}
    +
    \tfrac{1}{4} \bar{\Tilde{\mathfrak{h}}}^{T}{} \xi^{* \alpha }_s D_{\alpha }p
    +
    \frac{\!{}{}{}{}{}^{(3)}{}G^{\alpha \beta } \bar{\mathfrak{h}}^{*}_{s \alpha \beta } \bar{\Tilde{\mathfrak{h}}}^{T}{}}{64 \pi}
    - \frac{\!{}{}{}{}{}^{(3)}{}G^{\alpha \beta } \bar{\Tilde{\mathfrak{h}}}^{T}{}_{\alpha \beta } \bar{\mathfrak{h}}^{*}_s}{64 \pi} 
    \nonumber\\
    &
    +
    \tfrac{1}{16} \bar{\mathfrak{h}}_s^{*\alpha }{}_{\alpha } \bar{\Tilde{\mathfrak{h}}}^{T}{} h \rho 
    - \tfrac{1}{16} \bar{\Tilde{\mathfrak{h}}}^{T}{}^{\alpha }{}_{\alpha } \bar{\mathfrak{h}}^{*}_s h \rho 
    +
    \frac{D^{\alpha }\bar{\Tilde{\mathfrak{h}}}^{T}{} D_{\beta }\bar{\mathfrak{h}}^{*}_{s \alpha }{}^{\beta }}{64 \pi} -  \frac{D^{\alpha }\bar{\mathfrak{h}}^{*}_s D_{\beta }\bar{\Tilde{\mathfrak{h}}}^{T}{}_{\alpha }{}^{\beta }}{64 \pi}
    \bigg]
\end{align}
\end{widetext}
where $\Tilde{a}_s(\omega)$ are the Fourier components of the mode amplitudes. 
The right-hand side evaluates to
\begin{align}\label{eq:schematic-boundary-love}
    &-\oint_{R=R_{\star}} \mathcal{R}^{\alpha}_{\mathrm{Love}} n_{\alpha} (R+M)^2 d \Omega
    =
    \nonumber\\
    &
    -\oint_{R=R_{\star}} n^{\alpha} (R+M)^2 d\Omega 
    \bigg[ 
    \frac{\bar{\Tilde{\mathfrak{h}}}^{T}{} D_{\alpha }\bar{\mathfrak{h}}^{*}}{64 \pi} -  \frac{\bar{\mathfrak{h}}^{*} D_{\alpha }\bar{\Tilde{\mathfrak{h}}}^{T}{}}{64 \pi} 
    \nonumber\\
    &-  \frac{\bar{\Tilde{\mathfrak{h}}}^{T}{} D_{\sigma }\bar{\mathfrak{h}}^{*}_{\alpha }{}^{\sigma }}{64 \pi} + \frac{\bar{\mathfrak{h}}^{*} D_{\sigma }\bar{\Tilde{\mathfrak{h}}}^{T}{}_{\alpha }{}^{\sigma }}{64 \pi}
    \bigg] \nonumber\\
    &=
    \frac{4\pi}{2\ell+1} \tilde{I}_{\ell m} \tilde{d}_{\ell m}\mathcal{Q}(\ell,z_0,\omega)
    \,,
\end{align}
where $\mathcal{Q}(\ell, z_{0}, \omega)$ is an undetermined factor and $z_{0} = 1-2M/(R_{\mathrm{\star}}+M)$.
In the Newtonian limit $\mathcal{Q}_{\mathrm{Newt}} = 1$.
In GR, we do not have an explicit expression valid for all values of $\ell$ for $\mathcal{Q}$. However, for a given value of $\ell$, we can derive the expressions for $\mathcal{Q}$ by using the frequency-domain resummation introduced in~\cite{HegadeKR:2024agt}. 
We have derived the analytical solutions for $h_{\mu\nu}^{\mathrm{T}}$ using this method, and, from these solutions, we can obtain the expression for $\mathcal{Q}(\ell=2)$, namely
\begin{align}
    \mathcal{Q}(\ell=2, z_{0}, \omega) &= \frac{(3 z_0+1)^2}{4 \sqrt{z_0} (z_0+1)^2} + \mathcal{O}{(M\omega)^2}\,.
\end{align}
The $\mathcal{O}{(M\omega)^2}$ term is too long to be displayed here, so we provide explicit expressions in the supplementary {\texttt{MATHEMATICA}} file.

Combining Eqs.~\eqref{eq:LHS-for-love} and \eqref{eq:schematic-boundary-love} we see that
\begin{widetext}
\begin{align}\label{eq:multipole-GR-expr}
    &\sum_{s}  \tilde{a}_{s}(\omega)
    \int_{\Sigma_{\mathrm{star}}} \sqrt{\gamma} d^3 x \bigg[ 
    \tfrac{1}{4} h \xi^{*\alpha }_s \rho D_{\alpha }\bar{\Tilde{\mathfrak{h}}}^{T}{}
    +
    \tfrac{1}{4} \bar{\Tilde{\mathfrak{h}}}^{T}{} \xi^{* \alpha }_s D_{\alpha }p
    +
    \frac{\!{}{}{}{}{}^{(3)}{}G^{\alpha \beta } \bar{\mathfrak{h}}^{*}_{s \alpha \beta } \bar{\Tilde{\mathfrak{h}}}^{T}{}}{64 \pi}
    - \frac{\!{}{}{}{}{}^{(3)}{}G^{\alpha \beta } \bar{\Tilde{\mathfrak{h}}}^{T}{}_{\alpha \beta } \bar{\mathfrak{h}}^{*}_s}{64 \pi} 
    +
    \tfrac{1}{16} \bar{\mathfrak{h}}_s^{*\alpha }{}_{\alpha } \bar{\Tilde{\mathfrak{h}}}^{T}{} h \rho 
    - \tfrac{1}{16} \bar{\Tilde{\mathfrak{h}}}^{T}{}^{\alpha }{}_{\alpha } \bar{\mathfrak{h}}^{*}_s h \rho 
    \nonumber\\
    &+
    \frac{D^{\alpha }\bar{\Tilde{\mathfrak{h}}}^{T}{} D_{\beta }\bar{\mathfrak{h}}^{*}_{s \alpha }{}^{\beta }}{64 \pi} -  \frac{D^{\alpha }\bar{\mathfrak{h}}^{*}_s D_{\beta }\bar{\Tilde{\mathfrak{h}}}^{T}{}_{\alpha }{}^{\beta }}{64 \pi}
    \bigg]
    =
    \frac{4\pi}{2\ell + 1} \tilde{I}_{\ell m }(\omega) \tilde{d}_{\ell m}(\omega)
    \mathcal{Q}
    \,.
\end{align}
\end{widetext}
In the expression above $\tilde{h}^T$ is a solution that satisfies the Hamiltonian and momentum constraints and matches smoothly to the external tidal solution at the surface of the star.
In Newtonian gravity, $\bar{\tilde{h}}^T_{\mathrm{Newt}} = \frac{16 \pi}{2\ell + 1}r^{\ell} \tilde{d}_{\ell m}(\omega) Y_{\ell m}$, $h=1$, $\mathcal{Q}_{\mathrm{Newt}}=1$ and the contribution to the multipole moment is dominated by the first term in the above equation, leading to the familiar expression for the multipole moment in terms of mode amplitudes~\cite{Lai:1993di}
\begin{align}
    &\tilde{I}_{\ell m,\mathrm{Newt}}(\omega)
    =
    \nonumber\\
    &
    \left.
    \sum_{s} \tilde{a}_{s}(\omega) \int_{\Sigma_{\mathrm{star}}}
    \sqrt{\gamma} d^3 x \left[
    \xi^{*\alpha }_s \rho D_{\alpha }(r^{\ell} Y_{\ell m})
    \right]
    \right|_{\mathrm{Newt}}
    \,.
\end{align}

We can use Eq.~\eqref{eq:multipole-GR-expr} to obtain the expression for the dynamical tidal response presented in Eq.~\eqref{eq:dynamical-tidal-response-func-GR} of the main text.
First, we note that the right hand side of Eq.~\eqref{eq:multipole-GR-expr} can be simplified to
\begin{align*}
    \sum_{s}
    \frac{4\pi R^{\ell}_{\star}}{2 \ell + 1} M_{\star} \Tilde{d}_{\ell m}(\omega) \Tilde{a}_{s}(\omega) \mathcal{G}_{s}
\end{align*}
where
\begin{align}\label{eq:Gs-def}
    &\mathcal{G}_{s} = 
    \left(\frac{2\ell+1}{4\pi \tilde{d}_{\ell m}} \right)
    \frac{1}{M_{\star} R_{\star}^{{\ell}}}
    \int_{\Sigma_{\mathrm{star}}} \sqrt{\gamma} d^3 x \bigg[ 
    \tfrac{1}{4} h \xi^{*\alpha }_s \rho D_{\alpha }\bar{\Tilde{\mathfrak{h}}}^{T}{}
    \nonumber\\
    &+
    \tfrac{1}{4} \bar{\Tilde{\mathfrak{h}}}^{T}{} \xi^{* \alpha }_s D_{\alpha }p
    +
    \frac{\!{}{}{}{}{}^{(3)}{}G^{\alpha \beta } \bar{\mathfrak{h}}^{*}_{s \alpha \beta } \bar{\Tilde{\mathfrak{h}}}^{T}{}}{64 \pi}
    - \frac{\!{}{}{}{}{}^{(3)}{}G^{\alpha \beta } \bar{\Tilde{\mathfrak{h}}}^{T}{}_{\alpha \beta } \bar{\mathfrak{h}}^{*}_s}{64 \pi} 
    \nonumber\\
    &+
    \tfrac{1}{16} \bar{\mathfrak{h}}_s^{*\alpha }{}_{\alpha } \bar{\Tilde{\mathfrak{h}}}^{T}{} h \rho 
    - \tfrac{1}{16} \bar{\Tilde{\mathfrak{h}}}^{T}{}^{\alpha }{}_{\alpha } \bar{\mathfrak{h}}^{*}_s h \rho 
    \nonumber\\
    &+
    \frac{D^{\alpha }\bar{\Tilde{\mathfrak{h}}}^{T}{} D_{\beta }\bar{\mathfrak{h}}^{*}_{s \alpha }{}^{\beta }}{64 \pi} -  \frac{D^{\alpha }\bar{\mathfrak{h}}^{*}_s D_{\beta }\bar{\Tilde{\mathfrak{h}}}^{T}{}_{\alpha }{}^{\beta }}{64 \pi}
    \bigg]\,.
\end{align}
The left hand side of Eq.~\eqref{eq:multipole-GR-expr} can be simplified to
\begin{align*}
    \frac{8\pi}{2\ell+1} K_{\ell m} R_{\star}^{2\ell + 1} (\Tilde{d}_{\ell m})^2 \mathcal{Q}(\ell,z_0,\omega)
\end{align*}
using $\tilde{I}_{\ell m}(\omega) = 2 \tilde{K}_{\ell m}(\omega) R_{\star}^{2\ell + 1} \tilde{d}_{\ell m}(\omega)$.
Finally, we can use the solutions for the amplitudes provided in Eq.~\eqref{eq:sol-a-freq} of the main text to obtain
\begin{align}
    &\frac{8\pi}{2\ell+1} K_{\ell m} R_{\star}^{2\ell + 1} (\Tilde{d}_{\ell m})^2 \mathcal{Q}(\ell,z_0,\omega)
    \nonumber\\
    &=
    \sum_{s}
    \frac{4\pi R^{\ell}_{\star}}{2 \ell + 1} M_{\star} \Tilde{d}_{\ell m}(\omega) \Tilde{a}_{s}(\omega) \mathcal{G}_{s} 
    \nonumber\\
    &=
    \left(\frac{4\pi}{2 \ell + 1}\right)^2(\Tilde{d}_{\ell m})^2 R_{\star}^{2\ell+1}
    \sum_{s}
    \frac{I_{s} \mathcal{G}_{s}}{\left[1 - (\omega/\omega_s)^2\right] \mathscr{N}_s}
    \,.
\end{align}
Simplifying this equation results in Eq.~\eqref{eq:dynamical-tidal-response-func-GR} of the main text.
\subsection{Expressions for the tidal force, tidal stress-energy tensor and the dimensionless overlap integrals}
We here list the expressions appearing in the decomposition of the tidal force density $F^{\mu}$, the tidal stress-energy tensor $F_{\mu\nu}$ and the dimensionless multipole moments $I_{s}$ and $\mathcal{G}_{s}$.
The decomposition of the tidal force density is given by
\begin{subequations}\label{eq:Force-dens-expr}
\begin{align}
    &F_{\mu} = F_{\mu}^{\mathrm{dens,T}} + F_{\mu}^{\mathrm{T,grad.\,p}} + F_{\mu}^{\mathrm{dens,vec}} \,,\\
    &F_{\alpha}^{\mathrm{dens,T}} = \tfrac{1}{2} h \rho D_{\alpha }\mathfrak{h}^T \,,\\
    &F_{\alpha}^{\mathrm{T,grad.\,p}} =
    - \tfrac{1}{2} \mathfrak{h}^{T\beta }{}_{\beta } D_{\alpha }p -  \frac{\mathfrak{h}^{T\beta }{}_{\beta } p \Gamma D_{\alpha }p}{2 h \rho}
    \nonumber\\
    &\hspace{1cm} +
    \frac{1}{2} D_{\alpha} \left[\mathfrak{h}^{T\beta }{}_{\beta } p \Gamma  \right]
    ,\\
    &F_{\alpha}^{\mathrm{dens,vec}}
    =
    - \frac{h \rho \partial_t^{2} \mathfrak{h}^{T}_{\alpha }}{N} \,.
\end{align}
\end{subequations}
\begin{widetext}
The tidal stress-energy tensor is given by
\begin{align}\label{eq:tidal-stress-equation}
    &F_{\mu \nu}
    =
    -E_{\alpha \beta}^{\mathrm{grav}} \gamma^{\alpha}_{\mu} \gamma^{\beta}_{\nu} \nonumber\\
    &=
4 \pi \gamma_{\mu \nu } \mathfrak{h}^{T}{}^{\alpha }{}_{\alpha } p - 8 \pi \mathfrak{h}^{T}{}_{\mu \nu } p - 4 \pi \gamma_{\mu \nu } \mathfrak{h}^{T}{} p -  \tfrac{1}{2} \gamma_{\mu \nu } \mathfrak{h}^{T}{}^{\alpha \beta } \!^{(3)}{}R_{\alpha \beta } + \mathfrak{h}^{T}{}_{\nu }{}^{\alpha } \!^{(3)}{}R_{\mu \alpha } -  \tfrac{1}{2} \mathfrak{h}^{T}{}^{\alpha }{}_{\alpha } \!^{(3)}{}R_{\mu \nu } \nonumber \\ 
& + \tfrac{1}{2} \mathfrak{h}^{T}{} \!^{(3)}{}R_{\mu \nu } + \mathfrak{h}^{T}{}_{\mu }{}^{\alpha } \!^{(3)}{}R_{\nu \alpha } + \tfrac{1}{4} \gamma_{\mu \nu } \mathfrak{h}^{T}{}^{\alpha }{}_{\alpha } \!^{(3)}{}R -  \tfrac{1}{2} \mathfrak{h}^{T}{}_{\mu \nu } \!^{(3)}{}R -  \tfrac{1}{4} \gamma_{\mu \nu } \mathfrak{h}^{T}{} \!^{(3)}{}R - 4 \pi \gamma_{\mu \nu } \mathfrak{h}^{T}{}^{\alpha }{}_{\alpha } p \Gamma \nonumber \\ 
& + \frac{\mathfrak{h}^{T}{}_{\mu \nu } D_{\alpha }D^{\alpha }N}{N} + \frac{\gamma_{\mu \nu } \mathfrak{h}^{T}{} D_{\alpha }D^{\alpha }N}{2 N} -  \tfrac{1}{2} D_{\alpha }D_{\mu }\mathfrak{h}^{T}{}_{\nu }{}^{\alpha } -  \tfrac{1}{2} D_{\alpha }D_{\nu }\mathfrak{h}^{T}{}_{\mu }{}^{\alpha } + \frac{\gamma_{\mu \nu } D_{\alpha }N D^{\alpha }\mathfrak{h}^{T}{}}{N} \nonumber \\ 
& -  \frac{\gamma_{\mu \nu } D_{\alpha }\mathfrak{h}^{T}{}^{\beta }{}_{\beta } D^{\alpha }N}{2 N} + \frac{D_{\alpha }\mathfrak{h}^{T}{}_{\mu \nu } D^{\alpha }N}{2 N} + \frac{\gamma_{\mu \nu } D^{\alpha }N D_{\beta }\mathfrak{h}^{T}{}_{\alpha }{}^{\beta }}{N} + \tfrac{1}{2} \gamma_{\mu \nu } D_{\beta }D_{\alpha }\mathfrak{h}^{T}{}^{\alpha \beta } + \tfrac{1}{2} \gamma^{\alpha \beta } D_{\beta }D_{\alpha }\mathfrak{h}^{T}{}_{\mu \nu } \nonumber \\ 
& + \tfrac{1}{2} \gamma^{\alpha \beta } \gamma_{\mu \nu } D_{\beta }D_{\alpha }\mathfrak{h}^{T}{} + \frac{\gamma_{\mu \nu } \mathfrak{h}^{T}{}^{\alpha \beta } D_{\beta }D_{\alpha }N}{N} -  \frac{\gamma_{\mu \nu } \mathfrak{h}^{T}{}^{\alpha }{}_{\alpha } D_{\beta }D^{\beta }N}{2 N} -  \tfrac{1}{2} \gamma^{\alpha \beta } \gamma^{\gamma \delta } \gamma_{\mu \nu } D_{\delta }D_{\gamma }\mathfrak{h}^{T}{}_{\alpha \beta } \nonumber \\ 
& -  \frac{D^{\alpha }N D_{\mu }\mathfrak{h}^{T}{}_{\nu \alpha }}{2 N} -  \frac{\mathfrak{h}^{T}{}_{\nu }{}^{\alpha } D_{\mu }D_{\alpha }N}{N} -  \frac{D^{\alpha }N D_{\nu }\mathfrak{h}^{T}{}_{\mu \alpha }}{2 N} -  \frac{D_{\mu }N D_{\nu }\mathfrak{h}^{T}{}}{2 N} -  \frac{D_{\mu }\mathfrak{h}^{T}{} D_{\nu }N}{2 N} -  \frac{\mathfrak{h}^{T}{}_{\mu }{}^{\alpha } D_{\nu }D_{\alpha }N}{N} \nonumber \\ 
& + \tfrac{1}{2} D_{\nu }D_{\mu }\mathfrak{h}^{T}{}^{\alpha }{}_{\alpha } -  \tfrac{1}{2} D_{\nu }D_{\mu }\mathfrak{h}^{T}{} + \frac{\mathfrak{h}^{T}{}^{\alpha }{}_{\alpha } D_{\nu }D_{\mu }N}{2 N} -  \frac{\mathfrak{h}^{T}{} D_{\nu }D_{\mu }N}{2 N} + \frac{\gamma_{\mu \nu } \partial_t^{2} \mathfrak{h}^{T}{}^{\alpha }{}_{\alpha }}{2 N^2} -  \frac{\partial_t^{2} \mathfrak{h}^{T}{}_{\mu \nu }}{2 N^2} \nonumber \\ 
& -  \frac{\gamma_{\mu \nu } D^{\alpha }N \partial_t^{2} \mathfrak{h}^{T}{}_{\alpha }}{N^2} + \frac{D_{\nu }N \partial_t^{2} \mathfrak{h}^{T}{}_{\mu }}{2 N^2} + \frac{D_{\mu }N \partial_t^{2} \mathfrak{h}^{T}{}_{\nu }}{2 N^2} -  \frac{\gamma_{\mu \nu } \partial_t^{2} D_{\alpha }\mathfrak{h}^{T}{}^{\alpha }}{N} + \frac{\partial_t^{2} D_{\mu }\mathfrak{h}^{T}{}_{\nu }}{2 N} \nonumber \\ 
& + \frac{\partial_t^{2} D_{\nu }\mathfrak{h}^{T}{}_{\mu }}{2 N}
\,.
\end{align}
Using Eqs.~\eqref{eq:Force-dens-expr} and \eqref{eq:tidal-stress-equation} and \eqref{eq:Is-def}, one can show that
\begin{subequations}
\begin{align}
    &I_{s} = I_{s}^{\mathrm{dens,T}} + I_{s}^{\mathrm{T,grad.\,p}} + I_{s}^{\mathrm{dens,vec}}
    +
    I_{s}^{\mathrm{TS}}
    \,,\\
    &I_{s}^{\mathrm{dens,T}} =   
    \left(\frac{2\ell+1}{4\pi \tilde{d}_{\ell m}} \right)
    \frac{1}{M_{\star} R_{\star}^{{\ell}}}\int_{\Sigma_{\mathrm{star}}} d^3 x \sqrt{\gamma} N \xi^{*}_{{\beta},s} \Tilde{F}^{\beta}_{{\mathrm{dens,T}}}
    \,,\\
    &I_{s}^{\mathrm{T,grad.\,p}} = 
    \left(\frac{2\ell+1}{4\pi \tilde{d}_{\ell m}} \right)
    \frac{1}{M_{\star} R_{\star}^{{\ell}}}\int_{\Sigma_{\mathrm{star}}} d^3 x \sqrt{\gamma} N \xi^{*}_{{\beta},s} \Tilde{F}^{\beta}_{{\mathrm{T,grad.\,p}}}
    \,,\\
    &I_{s}^{\mathrm{dens,vec}} =
    \left(\frac{2\ell+1}{4\pi \tilde{d}_{\ell m}} \right)
    \frac{1}{M_{\star} R_{\star}^{{\ell}}}\int_{\Sigma_{\mathrm{star}}} d^3 x \sqrt{\gamma} N \xi^{*}_{{\beta},s} \Tilde{F}^{\beta}_{\mathrm{dens,vec}}
    \,,\\
    &I_{s}^{\mathrm{TS}} =
    \left(\frac{2\ell+1}{4\pi \tilde{d}_{\ell m}} \right)
    \frac{1}{M_{\star} R_{\star}^{{\ell}} (16 \pi)}\int_{\Sigma_{\mathrm{star}}} d^3 x \sqrt{\gamma} N h^{*}_{{\alpha \beta},s} \Tilde{F}^{\alpha \beta}
    \,.
\end{align}
\end{subequations}
Finally, from Eq.~\eqref{eq:Gs-def} we see that
\begin{subequations}
\begin{align}
    &\mathcal{G}_{s} = \mathcal{G}_{s}^{\mathrm{dens,T}} + \mathcal{G}_{s}^{\mathrm{T,grad.\,p}} + \mathcal{G}_{s}^{\mathrm{grav,T}}
    \,\\
    &\mathcal{G}_{s}^{\mathrm{dens,T}} = \left(\frac{2\ell+1}{4\pi \tilde{d}_{\ell m}} \right)
    \frac{1}{M_{\star} R_{\star}^{{\ell}}}\int_{\Sigma_{\mathrm{star}}} d^3 x \sqrt{\gamma} N \bigg[
    \tfrac{1}{4} h \xi^{*\alpha }_s \rho D_{\alpha }\bar{\Tilde{\mathfrak{h}}}^{T}{} 
    +
    \tfrac{1}{16} \bar{\mathfrak{h}}_s^{*\alpha }{}_{\alpha } \bar{\Tilde{\mathfrak{h}}}^{T}{} h \rho 
    - \tfrac{1}{16} \bar{\Tilde{\mathfrak{h}}}^{T}{}^{\alpha }{}_{\alpha } \bar{\mathfrak{h}}^{*}_s h \rho
    \bigg] \,,\\
    &\mathcal{G}_{s}^{\mathrm{T,grad.\,p}} =
    \left(\frac{2\ell+1}{4\pi \tilde{d}_{\ell m}} \right)
    \frac{1}{M_{\star} R_{\star}^{{\ell}}}\int_{\Sigma_{\mathrm{star}}} d^3 x \sqrt{\gamma} N \bigg[ 
    \tfrac{1}{4} \bar{\Tilde{\mathfrak{h}}}^{T}{} \xi^{* \alpha }_s D_{\alpha }p
    \bigg]
    \,,\\
    &\mathcal{G}_{s}^{\mathrm{grav,T}} =
    \left(\frac{2\ell+1}{4\pi \tilde{d}_{\ell m}} \right)
    \frac{1}{M_{\star} R_{\star}^{{\ell}}}\int_{\Sigma_{\mathrm{star}}} d^3 x \sqrt{\gamma} N 
    \bigg[ 
    \frac{\!{}{}{}{}{}^{(3)}{}G^{\alpha \beta } \bar{\mathfrak{h}}^{*}_{s \alpha \beta } \bar{\Tilde{\mathfrak{h}}}^{T}{}}{64 \pi}
    - \frac{\!{}{}{}{}{}^{(3)}{}G^{\alpha \beta } \bar{\Tilde{\mathfrak{h}}}^{T}{}_{\alpha \beta } \bar{\mathfrak{h}}^{*}_s}{64 \pi} 
    +
    \frac{D^{\alpha }\bar{\Tilde{\mathfrak{h}}}^{T}{} D_{\beta }\bar{\mathfrak{h}}^{*}_{s \alpha }{}^{\beta }}{64 \pi} -  \frac{D^{\alpha }\bar{\mathfrak{h}}^{*}_s D_{\beta }\bar{\Tilde{\mathfrak{h}}}^{T}{}_{\alpha }{}^{\beta }}{64 \pi}
    \bigg]
    \,.
\end{align}
\end{subequations}
\end{widetext}
\end{document}